\documentclass[12pt]{article}
\usepackage{authblk}
\usepackage{amsmath,amsfonts}
\usepackage{hyperref}
\usepackage{graphicx}
\usepackage[utf8]{inputenc}
\usepackage{slashed}
\usepackage{dsfont}
\usepackage{bm}
\usepackage{mathrsfs}
\usepackage{tensor}
\usepackage{xcolor}
\usepackage{amsthm}
\usepackage{amssymb}
\usepackage{epic}
\usepackage{eepic}
\usepackage[matrix,arrow]{xy}
\usepackage{epsfig}
\usepackage{url}
\usepackage{hyperref}
\usepackage{multirow}

\usepackage{psfrag}

\makeatletter\@addtoreset{equation}{section}\makeatother

\def\bC {\mathbb{C}}

\def\bR {\mathbb{R}}
\def\bZ {\mathbb{Z}}
\def\bN {\mathbb{N}}

\newcommand{\beq}{\begin{equation}}
\newcommand{\eeq}{\end{equation}}
\newcommand{\bea}{\begin{eqnarray}}
\newcommand{\eea}{\end{eqnarray}}

\newcommand{\vev}[1]{{\left< {#1} \right>}}

\newcommand{\nn}{\nonumber}

\newcommand{\mfr}{\mathfrak}

\newcommand{\cA}{{\mathcal A}}

\newcommand{\cC}{{\mathcal C}}
\newcommand{\cE}{{\mathcal E}}
\newcommand{\cD}{{\mathcal D}}
\newcommand{\cF}{{\mathcal F}}
\newcommand{\cG}{{\mathcal G}}

\newcommand{\cN}{{\mathcal N}}
\newcommand{\cO}{{\mathcal O}}

\newcommand{\cQ}{{\mathcal Q}}
\newcommand{\cR}{{\mathcal R}}

\usepackage{tikz}
\usetikzlibrary{automata}
\usetikzlibrary{arrows}
\usetikzlibrary{calc}
\usetikzlibrary{decorations.markings}
\usetikzlibrary{decorations.pathreplacing}
\usetikzlibrary{intersections}
\usetikzlibrary{positioning}
\usetikzlibrary{topaths}
\usetikzlibrary{shapes.geometric}
\usetikzlibrary{shapes.misc}
\tikzset{->-/.style = {
    decoration = {markings, mark = at position #1 with {\arrow{>}}},
    postaction = {decorate}}}
\tikzset{color-group/.style = {
    shape = circle,
    minimum size = 2.5ex,
    inner sep = .5ex,
    draw}}
\tikzset{flavor-group/.style = {
    shape = rectangle,
    minimum size = 2.5ex,
    inner sep = .5ex,
    draw}}
\tikzset{cf-group/.style = {
    shape = rounded rectangle,
    rounded rectangle right arc = none,
    draw}}
\tikzset{fc-group/.style = {
    shape = rounded rectangle,
    rounded rectangle left arc = none,
    draw}}
\tikzset{cross/.style={minimum width=1pt, path picture={
      \draw[black, very thick]
               (path picture bounding box.South east)
            -- (path picture bounding box.North west)
               (path picture bounding box.South west)
            -- (path picture bounding box.North east);
          }}}

\newcommand{\al}{\alpha}

\newcommand{\ga}{\gamma}
\newcommand{\Ga}{\Gamma}
\newcommand{\de}{\delta}
\newcommand{\De}{\Delta}
\newcommand{\vDe}{\varDelta}
\newcommand{\ep}{\epsilon}
\newcommand{\vep}{\varepsilon}
\newcommand{\tht}{\theta}

\newcommand{\la}{\lambda}

\newcommand{\om}{\omega}

\newcommand{\si}{\sigma}
\newcommand{\Si}{\Sigma}
\newcommand{\up}{\upsilon}

\newcommand{\ze}{\zeta}

\newcommand{\rf}[1]{(\ref{#1})}

\newcommand{\pa}{\partial}

\newcommand{\dd}{\mathrm{d}}
\newcommand{\lt}{\left}
\newcommand{\rt}{\right}
\newcommand{\lan}{\langle}
\newcommand{\ran}{\rangle}

\newcommand{\wt}{\widetilde}
\newcommand{\wh}{\widehat}

\newcommand{\ov}{\overline}

\newcommand{\mrm}{\mathrm}

\newcommand{\corr}[1]{\left\langle #1 \right\rangle}
\newcommand{\corrS}[1]{\left\langle #1 \right\rangle_{S^2}}
\newcommand{\corrR}[1]{\left\langle #1 \right\rangle_{\bR^2}}
\newcommand{\norm}[1]{\left\lVert #1 \right\rVert}

\newcommand{\ch}{\mathrm{c}}
\newcommand{\tc}{\mathrm{tc}}

\title{2D BPS Rings from Sphere Partition Functions}
\author[1,2]{Nafiz Ishtiaque}
\date{}
\affil[1]{\small Perimeter Institute for Theoretical Physics, Waterloo, ON N2L 2Y5, Canada}
\affil[2]{\small Department of Physics, University of Waterloo, Waterloo, ON N2L 3G1, Canada}
	
\begin{document}
	\maketitle
	
	\bibliographystyle{utphys}

	\abstract{We consider extremal correlation functions, involving arbitrary number of BPS (chiral or twisted chiral) operators and exactly one anti-BPS operator in 2D $\cN=(2,2)$ theories. These correlators define the structure constants in the rings generated by the BPS operators with their operator product expansions. We present a way of computing these correlators from the sphere partition function of a deformed theory using localization. Relating flat space and sphere correlators is nontrivial due to operator mixing on the sphere induced by conformal anomaly. We discuss the supergravitational source of this complication and a resolution thereof. Finally, we demonstrate the process for the Quintic GLSM and the Landau-Ginzburg minimal models.}
	
	\tableofcontents

\section{Prologue}
	Chiral and twisted chiral rings, collectively referred to as \emph{the BPS rings} in what follows, are algebraic structures that can be assigned to $\cN=(2,2)$ supersymmetric quantum field theories (QFTs) in two dimensions. On a general ground such assignments let us distinguish between different theories and identify various types of equivalence classes of QFTs and dualities. More specifically, the BPS rings are renormalization group (RG) invariants that can be used to distinguish between different universality classes of 2D $\cN=(2,2)$ theories.\footnote{These invariants consist of only local operators and they can not distinguish between theories with different non-local defects for example.} These rings are interesting objects from a mathematical point of view as well, as chiral and the twisted chiral rings of a given theory belong to two different topological sectors of the theory and their structures encode complex structure invariants and K\"ahler structure invariants of some geometric spaces associated to the theory \cite{Witten:1988xj,Witten:1991zz,Witten:1993yc}.
	\smallskip
	
	Analogous to the 2D case, chiral rings can be defined for 4D $\cN=2$ superconformal theories as well and recently a method has been used to compute the ring structure of these 4D chiral rings \cite{Gerchkovitz:2016gxx}, which entails computing the so called extremal correlation functions, using the exactly known sphere partition function of the theory \cite{Pestun:2007rz}. In this paper we use the same procedure to compute the BPS ring structure of 2D $\cN=(2,2)$ theories using the exactly known results regarding the 2D sphere partition functions \cite{Doroud:2012xw, Gomis:2012wy,  Doroud:2013pka,Benini:2012ui}. In order for us to use the sphere partition function and still be able to infer results for the theory on flat space, we require that we must be able to canonically place the flat space theory on a sphere. This forces us to restrict to 2D $\cN=(2,2)$ theories that flow to some conformal theories. One interesting feature of these rings in 2D is that, unlike their 4D analogues, they are not freely generated, and our procedure will generate the ring relations. This process does not rely on Mirror symmetry and therefore results obtained in this way can be used for independent checks of such symmetry.
	\smallskip
	
	The plan for the rest of the paper is as follows. In \S\ref{sec:chiralring} we establish the notations and conventions we use to characterize the BPS ring structures. In \S\ref{sec:CCR} we review, tailoring to the 2D case, the procedure put forward in \cite{Gerchkovitz:2016gxx} for computing chiral rings and finally in \S\ref{sec:examples} we apply this general procedure to compute the twisted chiral ring (consisting of the Coulomb branch operators) of the Quintic Calabi-Yau gauged linear sigma model (GLSM) and the chiral ring of the Landau-Ginzburg (LG) minimal models. In the appendices we present details about the superconformal algebra (\S\ref{sec:algebra}), supersymmetric backgrounds on the sphere (\S\ref{app:S2susy}), proof of a supersymmetric Ward identity we use (\S\ref{app:Ward}), and some explicit computations (\S\ref{sec:integral}).
	
	\textbf{Note:} After this paper was finished the paper \cite{Chen:2017fvl} came out with which this paper has a large overlap.

\section{The BPS Rings} \label{sec:chiralring}
	We first give the definition of the BPS ring in a superconformal theory, and then explain its definition for an ultraviolet (UV) theory with a conformal fixed point. We have included some details about the relevant $(2,2)$ superconformal algebra $su(2|2)$ in appendix \ref{sec:algebra}.
	\smallskip
	
	\subsubsection{In a superconformal theory} A \emph{superconformal primary} operator is one that is annihilated by all the $S$-supersymmetries:
	\beq
		\cO \mbox{ is a primary} \quad \Leftrightarrow \quad [S_\pm, \cO] = [\ov S_\pm, \cO] = 0\,. \label{def:primary}
	\eeq
	The anti-commutation relations of the $(2,2)$ superconformal algebra \rf{QQ} allow to consistently define the following types of primary operators with additional supersymmetry:
	\begin{subequations}\begin{align}
		\mbox{Chiral:} \qquad & [\ov Q_\pm, \cO] = 0\,, \label{chiraldef}\\
		\mbox{Anti-chiral:} \qquad & [Q_\pm, \cO] = 0\,, \\
		\mbox{Twisted chiral:} \qquad & [\ov Q_+, \cO] = [Q_-, \cO] = 0\,, \label{twchiraldef}\\
		\mbox{Twisted anti-chiral:} \qquad & [Q_+, \cO] = [\ov Q_-, \cO] = 0\,.
	\end{align}\label{ringdef}\end{subequations}
	Note that we use the name \emph{BPS} (\emph{anti-BPS}) to refer to both chiral and twisted chiral (anti-chiral and twisted anti-chiral). The above definitions apply to local and nonlocal operators alike but for this paper we are only concerned with local operators. Charges of chiral primaries under some of the generators of $su(2|2)$ are constrained: for example, using the $Q$-$S$ anti-commutators from \rf{QQ} it follows that the dimension and vector R-charge of a chiral primary $\cO$ are related, so are the dimension and the vector R-charge of an anti-chiral primary $\ov \cO$:
	\beq
		2\De(\cO) = J_V(\cO)\,, \qquad 2\De(\ov \cO) = -J_V(\ov\cO)\,. \label{DR}
	\eeq
	Such constraints lead to non-singular operator product expansion (OPE) between chiral primaries \cite{Lerche:1989uy}:
	\beq
		(\cO_1 \cO_2)(x) := \lim_{y \to x} \cO_1(x) \cO_2(y) =: \cO_3(x)\,,
	\eeq
	where $\cO_3$ is either zero or a chiral primary with dimension:
	\beq
		\De(\cO_3) = \De(\cO_1) + \De(\cO_2)\,.
	\eeq
	With this product the set of all chiral primaries becomes a ring called the \emph{chiral ring}, which we will denote as $\cR^\ch$. The twisted chiral ring, denoted $\cR^\tc$, is analogously defined as the ring of twisted chiral primaries.
	\smallskip
	
\subsubsection{Theories with conformal fixed points}
	The definition of primary opearators \rf{def:primary} does not apply in a non-conformal theory since the $S$-supersymmetries are not part of the symmetry in such case, so the BPS rings can not be defined in such a theory as the ring generated by the primaries. There is however an alternative definition of these rings which applies in this case. For that definition we need the following two nilpotent supercharges:
	\beq\begin{aligned}
		Q_A := \ov Q_+ + Q_-\,, & \qquad Q_B := \ov Q_+ + \ov Q_-\,, \\
		Q_A^2 = 0\,, & \qquad Q_B^2 = 0\,.
	\end{aligned}\eeq
	Now	the chiral (twisted chiral) ring can be defined as the $Q_B$-cohomology ($Q_A$-cohomology) of operators:
	\beq
		\cR^\ch := H_{Q_B}^\bullet\,, \qquad \cR^\tc := H_{Q_A}^\bullet\,, \label{Qcohdef}
	\eeq
	where the grading refers to the $U(1)_V$ R-charge for the chiral ring and the $U(1)_A$ R-charge for the twisted chiral ring.\footnote{$Q_A$ and $Q_B$ have charge $1$ under $U(1)_A$ and $U(1)_V$ respectively.}
	
	To see that these cohomologies define the same ring as the ring of chiral/twisted chiral primaries in a superconformal theory we need the following two observations:
	\begin{enumerate}
		\item Suppose $\cO$ is a $Q_B$-closed operator:
		\beq
			\lt[\ov Q_+ + \ov Q_-, \cO\rt] = 0\,. \label{QBclosed}
		\eeq
		If $\cO$ has spin\footnote{By spin we are referring to the charge for the generator $2J_L$ (where $J_L$ is the generator of rotation on $\bR^2$), note that $\ov Q_\pm$ has charge $\mp 1$ for this generator. (Details about the symmetry algebra are provided in \S\ref{sec:algebra}.)} $\al$ then rotating the above equation by an angle $\pi/2$ we get:
		\beq
			\lt[-i\ov Q_+ + i\ov Q_-,  e^{i \pi \al/2} \cO \rt] = 0 \label{QBclosedPi/2}
		\eeq
		Together, \rf{QBclosed} and \rf{QBclosedPi/2} imply:
		\beq
			\lt[\ov Q_+, \cO\rt] = \lt[\ov Q_-, \cO\rt] = 0\,.
		\eeq
		Thus we recover the chirality condition \rf{chiraldef}. Similarly it can be shown that being $Q_A$-closed is equivalent to being twisted chiral \rf{twchiraldef}.
		
		\item Any chiral (twisted chiral) operator is $Q_B$-cohomologous ($Q_A$-cohomologous) to a chiral (twisted chiral) primary \cite{Lerche:1989uy}. Furthermore, a $Q_{A/B}$-exact operator is not a primary, since a primary is defined as the operator in a superconformal multiplet with the lowest Weyl weight, whereas an operator $[Q_{A/B}, \cO]$ is in the same multiplet as $\cO$ while having a higher Weyl weight than $\cO$.
	\end{enumerate}
	The cohomological definition of the BPS rings \rf{Qcohdef} is perfectly sensible in the absence of conformal symmetry and coincides with the definition in terms of superconformal primaries at a conformal fixed point.

\subsubsection{Extremal correlators}
	Let us first define extremal correlators in an SCFT, and then explain why we can compute them in a UV theory with CFT fixed point.
	\smallskip
	
	Given two BPS operators, $\cO_1$ and $\cO_{2}$ of conformal dimensions $\De_1$ and $\De_2$ respectively, the field theory defines a Hermitian inner product:
	\beq
		\lan \cO_1, \cO_{2} \ran := \lim_{x \to \infty} |x|^{\De_1 + \De_2} \corrR{\cO_1(0) \ov \cO_{\ov 2}(x)} = \de_{\De_1, \De_2} \lim_{x \to \infty} |x|^{2\De_2} \corrR{\cO_1(0) \ov \cO_{\ov 2} (x)}\,, \label{inner}
	\eeq
	where $\ov \cO_{\ov 2}$ is the anti-BPS primary operator conjugate to $\cO_2$. The second equality follows from $U(1)_R$ selection rule\footnote{$U(1)_R = U(1)_V$ if the operators are chiral and $U(1)_R = U(1)_A$ if the operators are twisted chiral.} and the constraint \rf{DR} (and its analogue for the twisted case). In order to shorten the notation of \rf{inner} we define:
	\beq
		\ov\cO(\infty) := \lim_{x \to \infty} |x|^{2\De(\ov\cO)} \ov\cO(x)\,. \label{Oinf}
	\eeq
	The inner product \rf{inner} now becomes:
	\beq
		\lan \cO_1, \cO_{2} \ran = \corrR{\cO_1(0) \ov \cO_{\ov 2}(\infty)}\,. \label{inner2}
	\eeq
	The correlation functions of this form, i.e., with a BPS operator at $0$ and an anti-BPS operator at $\infty$, are called \emph{extremal correlators} on $\bR^2$.
	\smallskip
	
	There's a little more to the extremal correlators. Generally they are defined with an arbitrary number of BPS primaries $\cO_1, \cdots, \cO_m$ located at $x_1, \cdots, x_m$ respectively and one anti-BPS primary at infinity:
	\beq
		\corrR{\cO_1(x_1)\cdots \cO_m(x_m) \ov\cO(\infty)}\,,
	\eeq
	and this correlator is independent of the positions $x_1, \cdots, x_m$. We can see this by translating any of the BPS operators and using \rf{QQL-1}, for example, the infinitesimally translated correlator $\corrR{[L_{-1}, \cO_1](x_1)\cdots \cO_m(x_m) \ov\cO(\infty)}$ is proportional to:
	\beq
		\lim_{y \to \infty} |y|^{2\De(\cO)} \corrR{[\{Q_+, \ov Q_+\}, \cO_1](x_1)\cdots \cO_m(x_m) \ov\cO(y)} \,.
	\eeq
	Supersymmetric Ward identity allows us to pull $\ov Q_+$ out of $\cO_1$ and distribute it over the rest of the operators, all the BPS operators are annihilated by $\ov Q_+$ and when it acts on $\ov \cO$, the correlator behaves as $|y|^{-2\De(\cO) - 1}$ and the limit makes the contribution zero. This position independence of the extremal correlators allows us to bring all the chiral operators to one point (say at the origin).
	
	A similar argument shows that exact operators are zero inside extremal correlators and therefore the extremal correlators really define an inner product in the cohomology. Furthermore, the energy-momentum tensor couples to the linearized space-time metric via a D-term action \cite{Closset:2014pda}. Variation of a correlation function with respect to the metric then inserts an operator inside the correlator which is an integral over the entire superspace:
	\beq
		\de_{g_{\mu\nu}} \corrR{\cdots} \sim \corrR{\int \dd^2 \tht \dd^2 \ov\tht ( \cdots) \cdots}
	\eeq
	Such an integrated operator can be written as an exact operator \cite{Hori:740255} which implies that as long as $\corrR{\cdots}$ is an extremal correlator such variations vanish. This in particular implies that the extremal correlators are scale invariant, in other words, they are RG invariant and can be computed in a UV theory even when we are interested in an IR CFT fixed point.

\subsubsection{Basis, structure constants, norms and relations}
	In a finitely and freely generated ring\footnote{The 2D BPS rings are finitely but \emph{not} freely generated, we will discuss truncation by relations momentarily.} $\cR$ with a non-degenerate Hermitian inner product $\lan -, - \ran$, we can choose a minimal set of generators $\{\cO_1, \cO_2, \cdots, \cO_N\}$ and define a metric in their basis:
	\beq
		g_{i\ov j} := \lt\lan \cO_i, \cO_{j} \rt\ran\,. \label{def:metric}
	\eeq
	The inverse metric $g^{\ov i j}$ is defined by imposing:
	\beq
		g^{\ov i j} g_{j \ov k} = \de^{\ov i}_{\ov k}\,, \quad g_{i \ov j} g^{\ov j k} = \de_i^k\,.
	\eeq
	We define the ring structure by the structure constants in such a basis:
	\beq
		\cO_i \cO_j = \tensor{C}{_{ij}^k} \cO_k \quad \Leftrightarrow \quad \tensor{C}{_{ij}^k} = C_{ij \ov l} g^{\ov l k} \quad \mbox{where,} \quad C_{ij\ov l} := \lan \cO_i \cO_j, \cO_{l} \ran \label{def:structC}
	\eeq
	Furthermore, we can choose the basis in such way that the structure constants become trivial/diagonal in the following sense:\footnote{For two indices $i$ and $j$ referring to two operators $\cO_i$ and $\cO_j$, we use the index $i+j$ to refer to the operator with dimension equal to the sum of the dimensions of $\cO_i$ and $\cO_j$. For simplicity we are assuming that there is only one such operator, having more does not make any qualitative difference.}
	\beq	
		\tensor{C}{_{ij}^k} = \de_{i+j}^k\,. \label{diagonalC}
	\eeq
	Now all the nontrivial information about the ring structure is encoded in the norms of the basis vectors:
	\beq
		\norm{\cO_i} := \sqrt{\lt\lan \cO_i, \cO_{i} \rt\ran}\,.
	\eeq
	The constraint \rf{diagonalC} fixes the norms of all the basis vectors \emph{relative to each other}. To fix this arbitrariness in case of the BPS rings, we will fix the norm of the identity operator $\mathds 1$ to be $1$:\footnote{Note that $\norm{\mathds 1}^2 = \corrR{\mathds 1(x) \ov{\mathds 1}(\infty)} = Z$ where $Z$ is the partition function, therefore, in terms of Feynman diagrams, defining this norm to be one is equivalent to subtracting bubble diagrams from all our correlation functions.}
	\beq
		\lan \mathds 1, {\mathds 1} \ran := 1\,.
	\eeq
	
	Given a complete set of generators $\{\cO_1, \cdots, \cO_N\}$, a freely generated ring is simply the polynomial ring:
	\beq
		\cR = \bC[\cO_1, \cdots, \cO_N]\,.
	\eeq
	The only new addition to this discussion in the case of a ring \emph{with relations}, is that there will be some polynomials $p_a \in \bC[\cO_1, \cdots, \cO_N]$ for $a \in \{1, \cdots, M\}$ which will be identified with zero, i.e., we must impose the relations $p_a = 0$ for all $a \in \{1, \cdots, M\}$ and the ring will be given by:
	\beq
		\cR = \bC[\cO_1, \cdots, \cO_N]/\lan p_1, \cdots, p_M \ran\,.
	\eeq
	where $\lan p_1, \cdots, p_M \ran$ is the ideal generated by the polynomials $\{p_1, \cdots, p_M\}$.
	\smallskip
	
	In the context of the 2D $\cN=(2,2)$ BPS rings, the zero polynomials will appear as BPS operators with zero norm.\footnote{The identification of zero normed operators with identically zero operators is provided by the Reeh-Schlieder theorem \cite{2008arXiv0802.1854S}.} We will always choose a basis of the BPS operators with trivialized (as in \rf{diagonalC}) structure constants and the identity operator will be defined to have unit norm, therefore, according to the above discussion all the information of the BPS rings will be encoded in the extremal correlators $\corrR{\cO_i(0) \ov\cO_{\ov i}(\infty)}$, in particular, finding the relations will amount to finding BPS operators $\cO$ such that $\corrR{\cO(0) \ov\cO(\infty)} = 0$.

\section{Computing the Ring Structures} \label{sec:CCR}
	As explained in \S\ref{sec:chiralring}, a BPS ring structure is essentially defined by flat space extremal correlators $\lan \cO_1(0) \ov\cO_{\ov 2}(\infty)\ran_{\bR^2}$ of BPS primaries once a suitable basis has been chosen. A straightforward application of Weyl Ward identity tells us that if we put our theory on a sphere of radius $r$, then the extremal correlators on the sphere are related to the flat space correlators in the following way:
	\beq
		\corrR{\cO_1(0) \ov\cO_{\ov 2}(\infty)} = (2r)^{2\De(\cO_2)} \corrS{\cO_1(N) \ov\cO_{\ov 2}(S)}\,, \label{WeylWard}
	\eeq
	where $N$ and $S$ on the sphere are images of $0$ and $\infty$ on $\bR^2$ respectively, under an inverse stereographic projection. The $S^2$ correlators that appear in the above formula can be readily computed using localization. The main complication then, in using the above formula to compute the BPS ring structure constants, is that the identification between the flat space operators and the operators on the sphere is nontrivial due to operator mixing on the sphere. Mixing among operators of different dimensions can take place on the sphere because the sphere does not preserve scaling symmetry. Our task is therefore to ``unmix" the operators on the sphere and then use the Weyl Ward identity \rf{WeylWard} to compute the BPS ring structures. In this section we elaborate on this general procedure. We note that this process is essentially identical to the process of computing chiral rings in 4D $\cN=2$ SCFTs \cite{Gerchkovitz:2016gxx}.
	
\subsection{Extremal Correlators on $S^2$} \label{sec:exS2}

\subsubsection{Choice of a localizing supercharge} \label{sec:locQ}
	The first step in extracting the flat space extremal correlators from the sphere partition function is to compute their analogue on the sphere, such as $\lan \cO_i(N) \ov \cO_{\ov j}(S) \ran_{S^2}$,\footnote{North ($N$) and South ($S$) poles refer to two antipodal points on the sphere. We will take them to be $x=0$ and $x = \infty$ (in stereographic coordinate) for convenience.} using supersymmetric localization. We begin in this section by defining our choice of localizing supercharges for the two-sphere backgrounds described in \S\ref{app:S2susy} and some of their important properties:
	\begin{itemize}
		\item \textbf{Background-A:} In accordance with the notation of \S\ref{app:S2susy}, we define our choice of localizing supercharge by imposing the following chirality constraints on the constant Dirac spinors that parametrize the solutions of the Killing spinor equations (see \rf{solKSEA}):
		\beq
			\chi_{0-} = 0\,, \qquad \wt\chi_{0+} = 0\,. \label{constr}
		\eeq
		With these constraints the Killing spinors become chiral at the poles:
		\beq
			P_- \ep^A_{\chi_0, \wt\chi_0}(N) = P_+ \wt\ep^A_{\chi_0, \wt\chi_0}(N)  = 0\,, \quad
			P_+ \ep^A_{\chi_0, \wt\chi_0}(S) = P_- \wt\ep^A_{\chi_0, \wt\chi_0}(S)  = 0\,, \label{QA}
		\eeq
		where $P_\pm := \frac{1}{2}(1+\ga^3)$ are the chiral projectors. We will refer to this choice of supercharge as $\cQ_A$.
		\smallskip
		
		We recall that under a generic supercharge corresponding to a generic solution $\ep$ and $\wt\ep$ of the Killing spinor equations, a twisted chiral primary $Y$ and a twisted anti-chiral primary $\ov Y$, which are the bottom components of a twisted chiral mulitplet $(Y, \zeta, G)$ and a twisted anti-chiral  multiplet $(\ov Y, \ov \zeta, \ov G)$ respectively, transform as \rf{su21Atc}:
		\beq
			\de_{\ep, \wt\ep} Y(x) = \wt\ep_+(x) \ze_-(x) - \ep_-(x) \ze_+(x)\,, \qquad \de_{\ep, \wt\ep} \ov Y(x) = \wt\ep_-(x) \ze_+(x) - \ep_+(x) \ze_-(x)\,.
		\eeq
		Therefore, for the supercharge $\cQ_A$ corresponding to \rf{QA} we get:
		\beq
			\de_{\cQ_A} Y(N) = \de_{\cQ_A} \ov Y(S) = 0\,. \label{QAinv}
		\eeq
		This implies that insertions of twisted chiral and twisted anti-chiral primaries at the North and the South pole respectively are invariant under $\cQ_A$ and the corresponding correlators can be computed by supersymmetric localization using $\cQ_A$.\footnote{Such a supercharge was used in \cite{Doroud:2012xw} to compute the $su(2|1)_A$-invariant partition function using localization.}

		\item \textbf{Background-B:} We impose the same chirality constraints \rf{constr} on the constant spinors but this leads to different constraints for the Killing spinors of this background \rf{solKSEB}:
		\beq
			\wt\ep_{\chi_0, \wt\chi_0}^B(N) = 0\,, \qquad \ep_{\chi_0, \wt\chi_0}^B(S) = 0\,. \label{QB}
		\eeq
		We will refer to this choice of supercharge by $\cQ_B$.
		\smallskip
		
		We recall the transformations of a chiral primary $\phi$ and an anti-chiral primary $\ov\phi$, which are the bottom components of a chiral multiplet $(\phi, \psi, F)$ and an anti-chiral multiplet $(\ov\phi, \ov\psi, \ov F)$ respectively, under a generic supercharge \cite{Doroud:2012xw}:
		\beq
			\de_{\ep, \wt\ep} \phi(x) = \wt\ep(x) \psi(x)\,, \qquad \de_{\ep, \wt\ep} \ov\phi(x) = \ep(x) \ov\psi(x)\,.
		\eeq
		Therefore, according to \rf{QB} we have:
		\beq
			\de_{\cQ_B} \phi(N) = \de_{\cQ_B} \ov \phi(S) = 0\,, \label{QBinv}
		\eeq
		implying that we can compute correlators with insertions of chiral and anti-chiral primaries at the North and South pole respectively by supersymmetric localization using the supercharge $\cQ_B$.\footnote{Such a supercharge was used in \cite{Doroud:2013pka} to compute $su(2|1)_B$-invariant partition function and correlation functions of 2D gauge theories.}
	\end{itemize}

\subsubsection{A Ward identity and extremal correlators} \label{sec:Ward}
	A particularly convenient way to insert BPS (anti-BPS) primary operators at the North (South) pole of the sphere is to use a supersymmetric Ward identity. Before stating the identity, let us define for an arbitrary twisted chiral multiplet $\Psi = (Y, \ze, G)$ with a scalar bottom component:
	\beq
		\cG(\Psi) := G + \frac{\vDe(Y) - 1}{r} Y\,, \label{modG}
	\eeq
	where $\vDe(Y)$ denotes the Weyl weight (equal to the dimension for a scalar operator) of $Y$. Now we state the Ward identity:
	\begin{quote}
		Suppose we are given the following data in backgrounad-A: A supercharge $Q_A \in su(2|1)_A$, a $Q_A$-invariant operator\footnote{The operator $\cO$ does not have to be twisted chiral, it suffices that $\corrS{\cO}$ be an extremal correlator.} $\cO$ and a twisted chiral multiplet $\Psi = (Y, \ze, G)$ of arbitrary Weyl weight. Then, inside a correlator with $\cO$, the $su(2|1)_A$-invariant twisted F-term action for $\Psi$ localizes to the insertion of the bottom component $Y$ at the fixed point of $Q_A$ on the sphere (which we call the North pole $N$), in other words:
		\beq
			\corrS{\lt(\int_{S^2} \dd^2x \sqrt{g(x)}\, \cG(\Psi) \rt) \cO} = -4\pi r \corr{Y(N) \cO}\,, \label{wardtc}
		\eeq
		where $g$ is the determinant of the \emph{covariant} metric on the sphere. Similarly, the conjugate twisted F-term action of the twisted anti-chiral multiplet $\ov \Psi = (\ov Y, \ov \ze, \ov G)$ localizes to the insertion of the bottom component at the South pole (fixed point of $\ov Q_A$):
		\beq
			\corrS{\lt(\int_{S^2} \dd^2x \sqrt{g(x)}\, \cG(\ov\Psi) \rt) \cO} = 4\pi r \corr{\ov Y(S) \cO}\,. \label{wardtac}
		\eeq
	\end{quote}
	There is a parallel Ward identity for  background-B the statement of which simply replaces $Q_A$ with $Q_B$ and ``twisted chiral" with ``chiral". In \cite{Gerchkovitz:2014gta} this was proven for twisted chiral multiplets in background-A and chiral multiplets in background-B of Weyl weight 1.\footnote{Which results in the twisted F-term or the F-term action being a marginal deformation.} The proof for arbitrary Weyl weight requires only a trivial modification, we reproduce the modified proof in \S\ref{app:Ward} for reference.
	\smallskip
	
	The twisted F-terms or the F-terms can be used to deform the theory\footnote{We are assuming these deformation terms to be scalar so as not to break (Euclidean) Lorentz invariance. From now on we assume that all the non-trivial operators in the BPS rings are scalars, this will be true in the examples that we will consider.} in background-A or B respectively by introducing coupling constants of appropriate Weyl weights. For example, in background-A we can have the following deformation:\footnote{The normalization of the deformation term was chosen simply to cancel some numerical factors in \rf{wardtc} and \rf{wardtac}.}
	\beq
		S_A[X] \to S'_A[X; \tau, \ov \tau] := S_A[X] + \lt[-\frac{i\tau}{4\pi} \int_{S^2} \dd^2x \sqrt{g(x)}\, \cG(\Psi) + \mathrm{c.c.} \rt]\,, \label{def}
	\eeq
	where the bottom component $Y$ of the twisted chiral multiplet $\Psi = (Y, \ze, G)$ and the coupling constant $\tau$ have Weyl weights that satisfy:
	\beq
		\vDe(Y) + \vDe(\tau) = 1\,,
	\eeq
	and $X$ is merely a place-holder for all the dynamical fields. Using the Ward identities \rf{wardtc} and \rf{wardtac} we can now relate $\tau$-derivatives of the partition function to extremal correlators:
	\beq
		\frac{1}{Z_{S^2}^A} \frac{1}{ r^{m+n}} \pa_\tau^m \pa_{\ov\tau}^n Z_{S^2}^A(\tau, \ov\tau) \big|_{\tau, \ov\tau = 0} = \corrS{(iY)^m(N) (i\ov Y)^n(S)}\,, \label{exCorr}
	\eeq
	where $Z_{S^2}^A(\tau, \ov\tau)$ is the deformed partition function:\footnote{This deformed partition function does not need to be convergent, it is just a generating function with indeterminate variables $\tau$ and $\ov\tau$ for correlators with integrated operators, which, due to the Ward identities, become correlators with unintegrated twisted chiral and twisted anti-chiral primaries. We need only to be able to compute these correlation functions using localization.}
	\beq
		Z_{S^2}^A(\tau, \ov\tau) = \int \cD X\, e^{-S'_A[X; \tau, \ov \tau]}\,.
	\eeq
	We encode the equation \rf{exCorr} in the following correspondence between derivative with respect to a coupling, and the operator it inserts at a pole after localization:
	\beq
		\frac{1}{r}\pa_\tau \longleftrightarrow iY(N)\,, \qquad \frac{1}{r} \pa_{\ov\tau} \longleftrightarrow i\ov Y(S)\,. \label{dO}
	\eeq
	
	We can compute extremal correlators of chiral operators on the sphere similarly in  background-B.
	\smallskip
	
	\textbf{Remark:} If the undeformed action already contains a superpotential or twisted superpotential coupling then we can compute extremal correlators of the corresponding chiral or twisted chiral fields without any further deformation, just by taking derivatives with respect to the corresponding coupling constant. An example of this, which will be studied in detail later, is an abelian gauge theory in background-A where the action contains a complexified Fayet-Iliopoulos (FI) coupling $t \int_{S^2} \dd^2x \sqrt{g(x)}\, G_\Si$ where $G_\Si$ is the top component of a twisted chiral multiplet $\Si$ of Weyl weight\footnote{By the Weyl weight of a BPS multiplet we refer to the Weyl weight of its bottom component. In particular, by a twisted chiral multiplet $\Psi=(Y,\ze,G)$ of Weyl weight $w$ we mean that $Y$ has Weyl weight $w$. The Weyl weights of $\ze$ and $G$ are $\lt(w+\frac{1}{2}\rt)$ and $(w+1)$ respectively.} $1$ known as the field strength multiplet. The bottom component of this multiplet is a complex scalar $\si$ and we can therefore compute such extremal correlators as $\corrS{\si^m(N) \ov \si^n(S)}$ by evaluating derivatives of the partition function with respect to the FI parameters $t$ and $\ov t$ at arbitrary values of $t$ and $\ov t$. We will do this in \S\ref{sec:GLSM}.

\subsection{Chiral Ring Coefficients from Extremal Correlators on $S^2$} \label{sec:StoR}
	Knowing the extremal correlators on $S^2$, the next step is to extract from them the flat space extremal correlators.
	
\subsubsection{Operator mixing}
	As was pointed out in \cite{Gerchkovitz:2016gxx} for the case of 4D $\cN=2$ SCFTs, when put on a sphere, operators of different Weyl weights can mix due to the presence of scheme dependent Weyl symmetry breaking counterterms. This is true in two dimensions as well. The important difference between the two and four dimensional story is that, in four dimensions the $\cN=2$ supergravity background multiplet that goes into the counterterms causing the operator mixing had Weyl weight $2$, whereas the $\cN=(2,2)$ supergravity background multiplet in two dimensions responsible for operator mixing has Weyl weight $1$. This leads to the fact that in four dimensions two operators can mix on the sphere only if their Weyl weights differ by an even integer, on the other hand in two dimensions two operators with Weyl weights differing by any integer amount can mix. More specifically, on $S^2$, a chiral (twisted chiral) operator $\cO_w$ of Weyl weight $w$ can mix with all chiral (twisted chiral) operators of lower weights:\footnote{In the sum we are restricting to lower weights to avoid repeated counting, since mixing with an operator of higher weight is already considered as a mixing of the higher weighted operator with the lower weighted operator. Also, we are assuming that there is at most one operator with a given Weyl weight for simplicity. If there are more than one operators of a given Weyl weight then we only need to choose an ordering of these operators and all the computations follow without any qualitative modification.}
	\beq
		\cO_w \to \cO_w + \sum_{\substack{n \in \bN \\ 0 < n \le w}} \al_n(\tau_\mrm{mar}) r^{-n} \cO_{w-n}\,, \label{mixing}
	\eeq
	where the mixing coefficients $\al_n$ are arbitrary holomorphic functions of all the exactly marginal couplings, schematically written as $\tau_\mrm{mar}$. We now construct the $\cN=(2,2)$ supergravity counterterms giving rise to such mixings.
	
	There are two minimal versions of $\cN=(2,2)$ supergravity that differ in the choice of $U(1)$ R-symmetry that is gauged \cite{1987CQGra...4...11H, Gates:1995du, Grisaru:1995dr, Grisaru:1994dm, Closset:2014pda, Bae:2015eoa}. After choosing appropriate background values for the fields, these two versions reduce to background-A and background-B on $S^2$ preserving the vector and the axial R-symmetry respectively. Let us focus on the supergravity leading to background-A.
	
	We discuss the mixing of the bottom component of a twisted chiral multiplet $\wh \cO_w = (\cO_w, \ze_{\cO_w}, G_{\cO_w})$ of Weyl weight $w$. In order to compute correlation functions of the operator $\cO_w$ using the Ward identity \rf{wardtc} we need to deform the action, as in \rf{def}, by introducing a coupling. The manifestly supersymmetric way of doing this is to use superspace integrals to write the deformation terms. To that end we need to promote the coupling, which we denote as $\tau_{1-w}$ (making the Weyl weight explicit), to the bottom component of a background twisted chiral multiplet $\wh \tau_{1-w} = (\tau_{1-w}, \ze_{\tau_{1-w}}, G_{\tau_{1-w}})$. For this background multiplet to be supersymmetric, the $su(2|1)_A$ variations of the component fields must vanish. Consulting \rf{su21Atc} we find the following background values for the fermion and the top component (given the constant value of the bottom component):
	\beq
		\ze_{\tau_{1-w}} = 0\,, \qquad G_{\tau_{1-w}} = \frac{w-1}{r} \tau_{1-w}\,.
	\eeq
	Now the superspace integral representation of the deformation \rf{def} becomes:\footnote{The equality in \rf{twFDe} can be proven as follows. For a twisted chiral multiplet $\Psi_1 = (Y, \ze, G)$ of Weyl weight $w=1$, the relevant superspace integral just picks up the top component, i.e., $\int_{S^2} \dd^2x \int \dd^2 \wt\tht\, \cE_\tc\, \Psi_1 = \int_{S^2} \dd^2x \sqrt{g(x)}\, G$. That this is supersymmetric can also be checked by noting that the $su(2|1)_A$ variation of $G$, namely $\de G = \nabla_m(-i \wt\ep_- \ga^m \ze_- + i \ep_+ \ga^m \ze_+)$ (see \rf{su21Atc}), is a total derivative. Now assume $\Psi_w = (Y, \ze, G)$ is a twisted chiral multiplet of some arbitrary Weyl weight $w$ and $\wh \tau_{1-w} = \lt(\tau, 0, \frac{w-1}{r} \tau\rt)$ is a \emph{supersymmetric background} twisted chiral multiplet of Weyl weight $(1-w)$. Then $\wh\tau_{1-w} \Psi_w = \lt(\tau Y, \tau \ze, \tau \lt(G + \frac{w-1}{r}Y \rt) \rt)$ \cite{Closset:2014pda} is a twisted chiral multiplet of Weyl weight 1 and therefore $\int_{S^2} \dd^2x \int \dd^2 \wt\tht\, \cE_\tc\, \wh\tau_{1-w} \Psi_w = \int_{S^2} \dd^2x \sqrt{g(x)} \lt(\tau \lt(G + \frac{w-1}{r}Y \rt) \rt)$, and it is supersymmetric.}
	\beq
		-\frac{i\tau_{1-w}}{4\pi} I_{w,0} := -\frac{i}{4\pi} \int_{S^2} \dd^2x \int \dd^2 \wt\tht\, \cE_\tc\, \wh \tau_{1-w} \wh \cO_w = -\frac{i\tau_{1-w}}{4\pi} \int_{S^2} \dd^2x \sqrt{g(x)}\, \cG(\wh\cO_w)\,. \label{twFDe}
	\eeq
	where, as in the definition \rf{modG}, $\cG(\Psi) = G + \frac{w-1}{r} Y$. The supergravity counterterm that leads to the mixing of the operator $\cO_w$ with another twisted chiral operator $\cO_{w-n}$ of lower weight ($n \in \bN_{>0}$) necessarily involves the background coupling multiplet $\wh\tau_{1-w}$, the twisted chiral multiplet $\wh\cO_{w-n} := (\cO_{w-n}, \ze_{\cO_{w-n}}, G_{\cO_{w-n}})$ of weight $(w-n)$ and a background twisted chiral multiplet $\wh M = (M, \ze_R, -\cR/2)$ whose bottom component is a complex scalar of Weyl weight $1$ coming from the supergravity multiplet and whose top component is proportional to the scalar curvature of the space-time (this multiplet appeared in \cite{Ketov:1996es} in the context of 2D supergravity and in \cite{Gerchkovitz:2014gta} in constructing supergravity counterterms responsible for K\"ahler ambiguity in two-sphere partition function). On the sphere background, the scalar curvature is $\cR = 2/r^2$. As we did for the background coupling multiplet $\wh\tau_{1-w}$, we now find the supersymmetric background values for the component fields of $\wh M$ (this time given the constant value of the top component):
	\beq
		M = \frac{1}{r}\,, \qquad \ze_R = 0\,, \qquad -\frac{\cR}{2} = -\frac{1}{r^2}\,.
	\eeq
	Apart from the multiplets just mentioned, we have the freedom to include an arbitrary holomorphic function $\al$ of the exactly marginal couplings $\tau_\mrm{mar}$, including this we can now write down the mixing counterterm:
	\beq
		-\frac{i\tau_{1-w}}{4\pi} I_{w, n} := -\frac{i}{4\pi}\int_{S_2} \dd^2x \int \dd^2 \wt \tht\, \cE_\tc\, \wh \tau_{1-w} \al(\wh\tau_\mrm{mar}) \wh M^n \wh \cO_{w-n}\,, \label{mixSupInt}
	\eeq
	where we have promoted the exactly marginal couplings to background twisted chiral multiplets of Weyl weight $0$. Just to avoid cluttering the notation too much, let us introduce a symbol for the product multiplet:
	\beq
		\wh\cO_{w,n}^\al := \al(\wh \tau_\mrm{mar}) \wh M^n \wh \cO_{w-n}\,.
	\eeq
	Since this is a multiplet of Weyl weight $w$, we can use \rf{twFDe} to evaluate the superspace integral in \rf{mixSupInt} which leads to:
	\beq
		I_{w,n} = \int_{S^2} \dd^2x \sqrt{g(x)}\, \cG(\wh\cO_{w,n}^\al)\,. \label{mixcounter}
	\eeq
	The Ward identity \rf{wardtc} tells us that, inside an extremal correlator, the integrated operator $I_{w,n}$ will localize to the insertion of the bottom component of $\wh\cO_{w,n}^\al$ at the North pole. The bottom component of a product multiplet is simply the product of the bottom components of the individual multiplets in the product \cite{Closset:2014pda}. Therefore, in presence of the counterterm \rf{mixcounter}, the correspondence between coupling derivatives and operators \rf{dO} is modified:
	\beq
		\frac{1}{r} \pa_{\tau_{1-w}} \longleftrightarrow i\cO_w(N) + i\al(\tau_\mrm{mar}) r^{-n} \cO_{w-n}(N)\,. \label{dO2}
	\eeq
	In general, we must consider all possible counterterms, $\tau_{1-w} I_{w,n}$ for all $n \in \bN$ with $0 < n \le w$ and this leads to the general form of the mixing \rf{mixing}.

\subsubsection{``Un-mixing" the operators}
	Let us define $\mfr O_w$ to be the mixed operator in \rf{mixing}:
	\beq
		\mfr O_w := \cO_w + \sum_{\substack{n \in \bN \\ 0 < n \le w}} \al_n(\tau_\mrm{mar}) r^{-n} \cO_{w-n} \quad \Rightarrow \quad
		 \frac{1}{r} \pa_{\tau_{1-w}} \longleftrightarrow i \mfr O_w\,. \label{newO}
	\eeq
	Note that the mixing coefficients are scheme dependent,\footnote{A choice of scheme is a choice of the holomorphic functions $\al_n$ of the the exactly marginal couplings.} so these operators are not physical. But due to the mixing counterterms, such as \rf{mixcounter}, taking derivatives of the deformed sphere partition function with respect to the coupling constants computes extremal correlation functions of these operators:
	\beq
		\frac{1}{Z_{S^2}^A} \frac{1}{r^2} \pa_{\tau_{1-w}} \pa_{\ov\tau_{1-w'}} Z_{S^2}^A \big|_{\tau_{1-w} = \tau_{1-w'}=0} = \corrS{i\mfr O_w(N) i\ov{\mfr O}_{w'}(S)}\,. \label{newOcorr}
	\eeq
	We are of course interested in the flat space correlation functions of the physical operators, such as $\corrR{\cO_w(0) \ov\cO_{w'}(\infty)}$. Once we properly identify the flat space operators with their counterparts on the sphere, we can relate the correlators on $\bR^2$ with the correlatros on $S^2$ by the Weyl Ward identity \rf{WeylWard}.
	
	On flat space, operators of different Weyl weights are orthogonal, this changes on the sphere.\footnote{For example, if $\tau$ is an exactly marginal coupling then the partition function $Z_{S^2}(\tau, \ov\tau)$ depends on it and has a nonzero derivative: $\frac{1}{Z_{S^2}(\tau, \ov\tau)} \pa_\tau  Z_{S^2}(\tau, \ov\tau) = \corrS{\cO_\tau}$. Here $\cO_\tau$ is the bottom component of a BPS multiplet whose top component is an exactly marginal operator of Weyl weight $2$. The bottom component $\cO_\tau$ has Weyl weight $1$ and the fact that it has a nonzero one-point function indicates that it has mixed with the identity operator (of Weyl weight $0$).} It is a standard procedure to compute the inner products in an orthogonal basis (the $\cO_w$'s) given the inner products in the mixed basis (the $\mfr O_w$'s), called the \emph{Gram-Schmidt} procedure. In order to state the result, it is convenient to define some matrices. Given a complete set of operators $\{\mfr O_w\}$ indexed by their Weyl weights, define the following matrices:
	\beq
		M_{(w)} := \lt(\begin{array}{ccc} M_{0,0} & \cdots & M_{0,w} \\ \vdots & \ddots & \vdots \\ M_{w,0} & \cdots & M_{w,w} \end{array} \rt)\,, \qquad M_{i,j} := \corrS{i\mfr O_i(N) i\ov{\mfr O}_j(S) }\,. \label{Matrix}
	\eeq
	Now, we can express the flat space correlators of interest as follows (for $w \ge w'$):
	\beq
		\corrR{i\cO_w(0) i\ov\cO_{w'}(\infty)} = \de_{w, w'} (2r)^{2w'} \frac{\det M_{(w)}}{\det M_{(w-1)}}\,. \label{formula}
	\eeq
	We will use this formula in examples to compute chiral and twisted chiral ring relations in the following section.

\section{Some Examples} \label{sec:examples}
	In this section we illustrate the general points made so far by applying them to a couple of well known $\cN=(2,2)$ theories, namely the Quintic GLSM and Landau-Ginzburg minimal models.

\subsection{Twisted Chiral Ring of the Quintic GLSM} \label{sec:GLSM}
	This is a $U(1)$ gauge theory with $\cN=(2,2)$ supersymmetry and the bosonic global symmetry is $U(1)_L \times U(1)_V \times U(1)_A$, where $U(1)_L$ is the spacetime rotation and $U(1)_V, U(1)_A$ are the vector and axial R-symmetries respectively. It is a theory of six chiral multiplets $\Phi_i$ with $i \in \{1, \cdots, 6\}$  interacting via a superpotential:
	\beq
		W(\Phi_1, \cdots, \Phi_6) = \Phi_6 P(\Phi_1, \cdots, \Phi_5)\,,
	\eeq
	where $P$ is a homogeneous polynomial of degree five. We will denote the vector multiplet by $V$ and the associated \emph{twisted chiral} ``field strength" multiplet is defined as (in superfield notation):
	\beq
		\Si := \ov D_+ D_- V\,,
	\eeq
	where $\ov D_+$ and $D_-$ are two of the four superspace derivatives that commute with the supercharges. $\Si$ has Weyl weight\footnote{The kinetic term for the vector multiplet is normalized as $\frac{1}{e^2} \int \dd^2x \dd^2 \tht \dd^2 \ov \tht\, \Si \ov \Si$ where $e$ is the gauge coupling of dimension $1$.} $1$ and its top component is $\mrm D - iF_{12}$, where $\mrm D$ is the real scalar in $V$ and $F_{12}$ is the field strength of the gauge field in $V$. This field strength multiplet defines the twisted superpotential action:\footnote{Our normalization of this term has an extra factor of $\frac{1}{2\pi}$ compared to that of \cite{Doroud:2012xw, Benini:2012ui}.}
	\beq
		-\frac{t}{4\pi} \int_{\bR^2} \dd^2x \int \dd \tht^+ \dd \ov \tht^-\, \Si - \mrm{c.c.} = -\frac{i}{2\pi} \int_{S^2} \dd^2x \lt(\xi \mrm D - \frac{\tht}{2\pi} F_{12} \rt)\,, \label{twFS}
	\eeq
	where $\xi$ and $\tht$ are the real FI parameter and the topological theta angle respectively:
	\beq
		t = i \xi + \frac{\tht}{2\pi}\,. \label{FI}
	\eeq
	
	Charges of the fields of this theory under the gauge and the global symmetries are as follows (we have also included the charges of the superspace coordinates for quick reference):\footnote{The charges of the fields follow from the following arguments. The chiral fields appear in the action in the term $\int \dd^2x \int \dd \tht^+ \dd \tht^- \Phi_6 P(\Phi_1, \cdots, \Phi_5)$ and the twisted chiral field appears in $\int \dd^2x \int \dd \tht^+ \ov\tht^- \Si$ and these terms have to be neutral. The bosonic measure is neutral under all the symmetry groups. Noting that $\tht$ and $\dd\tht$ have opposite charges we see that the fermionic measure $\dd \tht^+ \dd \tht^-$ has charge $(0,0,-2,0)$ under $U(1)_L \times U(1)_\mathrm{gauge} \times U(1)_V \times U(1)_\cA$, and $\dd \tht^+ \dd \ov \tht^-$ has charge $(0,0,0,-2)$ under the same group.}
	\beq\begin{array}{c||c|c|c||c|c}
		& \Phi_1, \cdots, \Phi_5 & \Phi_6 & \hspace{.2cm} \Si \hspace{.2cm} & \hspace{.2cm} \tht^\pm \hspace{.2cm} & \hspace{.2cm} \ov\tht^\pm \hspace{.2cm} \\
		\hline
		U(1)_L & 0 & 0 & 0 & \pm & \pm\\
		U(1)_\mathrm{gauge} & 1 & -5 & 0 & 0 & 0\\
		U(1)_V & 2q_V & 2-10q_V & 0 & + & -\\
		U(1)_A & q_A & -5q_A & 2 & \pm & \mp\\
	\end{array}\eeq
	This model gets its name from the fact that for the value of the real FI parameter in a certain range (namely $\xi \gg 0$), the IR CFT fixed point of this theory is described by a non-linear sigma model with target \cite{Witten:1993yc}:
	\beq
		\Si = \Phi_6 = 0\,, \qquad \lt\{\lt(\sum_{i=1}^5 |\Phi_i|^2 = \Im t\rt) \rt\} \bigg / U(1)_\mathrm{gague} \cap \lt\{P(\Phi_1, \cdots, \Phi_5) = 0\rt\}\,,
	\eeq
	which is known as the Quintic Calabi-Yau (CY) threefold.\footnote{At the conformal fixed point the target space metric (i.e., the metric that appears in the kinetic term of the non-linear sigma model) is necessarily the Ricci flat metric.}
	\smallskip
	
\subsubsection{Twisted chiral ring}
	The twisted chiral ring $\cR^\tc$ of this theory is generated by the complex scalar operator $\si$, which is the bottom compotent of the field strength multiplet $\Si$.\footnote{It should be noted that the non linear sigma model description in the IR Calabi-Yau phase of this GLSM contains more twisted chiral operators which are fermionic. These are the operators that survive the A-twist \cite{Witten:1991zz}. However, the description of these operators in the GLSM is unclear. We thank Cyril Closset for pointing out this subtlety.} An orthogonal spanning set for $\cR^\tc$ is given by $\{\si^m\}_{m=0}^\infty$. This set also satisfies the triviality constraint for the structure constants \rf{diagonalC} since we have:
	\beq
		\si^m \si^n = \si^{m+n}\,.\label{diagonalCQuintic} 
	\eeq
	Therefore, after fixing the norm of $\si^0 = \mathds 1$ to be $1$, all we have left to compute to determine $\cR^\tc$ are the following extremal correlators:
	\beq
		\norm{\si^m}^2 = \corrR{\si^m(0) \ov \si^m(\infty)} \quad \forall m \ge 0\,.
	\eeq
	In the following we will compute these correlation functions and we will find that $\cR^\tc$ is not freely generated, we will find the null operators as well.

\subsubsection{From sphere to flat space}
	Extremal correlators of twisted chiral operators on the sphere can be computed by putting the theory in background-A (see \S\ref{sec:exS2}). The partition function of a generic $\cN=(2,2)$ gauge theory in background-A has been computed explicitly in \cite{Doroud:2012xw,Benini:2012ui} and this result has been applied to the specific case of the Quintic GLSM (among several others) in \cite{Jockers:2012dk}.\footnote{These localization results were further extended to extremal correlators on $S^2$ and used to find evidence for Seiberg-like dualities for $(2,2)$ gauge theories \cite{Benini:2012ui,Doroud:2012xw,Benini:2014mia}.} 
	\smallskip
	
	In the computation of the partition function in background-A, we can ignore the superpotential and only the twisted superpotential is important. The twisted superpotential action \rf{twFS} of the theory gets modified in the sphere background by the appearance of nontrivial integration measures:
	\beq
		-\frac{1}{4\pi} \int_{S^2} \dd^2x \int \dd^2 \wt\tht\, \cE_\tc\, \wh t\, \Si - \mathrm{c.c.} = -\frac{t}{4\pi} \int_{S^2} \dd^2x \sqrt{g(x)}\, \cG(\Si) - \mrm{c.c.} \,,
	\eeq
	This being the only twisted superpotential of the theory, the $su(2|1)_A$ preserving partition function $Z_{S^2}^A(t, \ov t)$ for the Quintic is a function only of the complexified FI parameter. Due to the presence of mixing counterterms (see \rf{mixcounter}), derivatives of the partition function $Z_{S^2}^A(t, \ov t)$ compute correlation functions of mixed operators. We denote these mixed operators as $\mfr s_m := \si^m + \cO(r^{-1})$ (c.f. \rf{newO}) so that $t$ and $\ov t$-derivatives of $Z_{S^2}^A$ can be equated readily with correlation functions of $\mfr s_m$ and $\ov{\mfr s}_n$ (c.f. \rf{newOcorr}):
	\beq
		\frac{1}{r^{m+n} Z_{S^2}^A} \pa_t^m \pa_{\ov t}^n Z_{S^2}^A(t, \ov t) = \corrS{\mfr s_m(N) \ov{\mfr s}_n(S)}\,. \label{newscorr}
	\eeq
	According to \rf{formula}, in terms of these correlators, the extremal correlators on the flat space are given by:
	\beq
		\corrR{\si^m(0) \ov\si^n(\infty)} = \de_{m,n} \frac{(2r)^{2n}}{Z_{S^2}^A} \frac{\det_{i,j\in\{0,\cdots, m\}} \pa^i_t \pa^j_{\ov t} Z_{S^2}^A}{\det_{i,j\in\{0,\cdots, m-1\}} \pa^i_t \pa^j_{\ov t} Z_{S^2}^A}\,. \label{key}
	\eeq
	From now on we will set the radius of the sphere to $1$ for simplicity.
	\smallskip

	The partition function $Z_{S^2}^A(t, \ov t)$ does not depend on the axial R-charges, so instead of writing $q_V$ all the time we will simply write $q$. For $\xi \gg 0$, the partition function can be written as \cite{Jockers:2012dk}:
	\beq
		Z_{S^2}^A(t,\ov t) = (w \ov w)^q \oint \frac{\dd \ep}{2\pi i} (w \ov w)^{-\ep} \frac{\pi^4 \sin(5\pi \ep)}{\sin^5(\pi \ep)} \lt| \sum_{k=0}^\infty (-w)^k \frac{\Ga(1+5k-5\ep)}{\Ga(1+k-\ep)^5} \rt|^2 \,, \label{ZA}
	\eeq
	where we have defined:\footnote{There is again a factor of $2\pi$ offset with respect to the convention of \cite{Jockers:2012dk}, where they use $e^{2\pi it}$. This offset compensates for our choice of normalization in \rf{twFS}.}
	\beq
		w := e^{i t}\,. \label{FIfug}
	\eeq
	In \rf{ZA}, the contour surrounds only the pole at $\ep = 0$ and in computing the absolute value of the infinite sum complex conjugation does not act on $\ep$. The infinite sum appearing in \rf{ZA} converges at $\ep = 0$ for large enough $\xi$, as can be seen from ratio test:
	\beq
		\la_k := (-w)^k \frac{\Ga(1+5k)}{\Ga(1+k)^5}\,, \qquad \frac{\la_{k+1}}{\la_k} \xrightarrow{k \to \infty} -e^{\frac{i\tht}{2\pi}-\xi} 5^5 \xrightarrow{\xi \to \infty} 0\,.
	\eeq
	Let us denote the series at $\ep=0$ as:
	\beq
		X( t) := \sum_{k=0}^\infty (-e^{i  t})^k \frac{\Ga(1+5k)}{\Ga(1+k)^5}\,. \label{X}
	\eeq
	The residue of the integrand at the pole is:
	\beq
		\underset{\ep=0}{\mathrm{Res}}\, \frac{(w \ov w)^{-\ep}}{2\pi i}  \frac{\pi^4 \sin(5\pi \ep)}{\sin^5(\pi \ep)} \lt| \sum_{k=0}^\infty (-w)^k \frac{\Ga(1+5k-5\ep)}{\Ga(1+k-\ep)^5} \rt|^2 = -\frac{10}{3\pi}i  \xi (-5\pi^2 + \xi^2) X( t) \ov X(\ov  t)\,.
	\eeq
	Therefore the partition function as a function of the FI parameter looks like:
	\beq
		Z_{S^2}^A( t, \ov t) = \frac{20}{3} e^{-2 q \xi } \xi (-5\pi^2 + \xi^2) X( t) \ov X(\ov  t)\,.
	\eeq
	One interpretation of this partition function is that it computes the K\"ahler potential of the moduli space of CFTs that can be reached from the GLSM under RG flow by varying the FI parameter \cite{Gerchkovitz:2014gta}:
	\beq
		Z_{S^2}^A(t, \ov t) = e^{-K(t,\ov t)}\,.
	\eeq
	From this perspective, the partition function is defined only upto a K\"ahler transformation of the K\"ahler potential:
	\beq
		K(t, \ov t) \to K(t, \ov t) + F(t) + \ov F(\ov t)\,,
	\eeq
	where $F$ and $\ov F$ are arbitrary holomorphic and anti-holomorphic functions (in particular, they can be taken as $F = \log X$, $\ov F = \log \ov X$), and a K\"ahler transformation can be interpreted as a change of the UV regularization scheme \cite{Gerchkovitz:2014gta}, which does not affect any physical observables.\footnote{A careful proof of the fact that a K\"ahler transformation does not affect the extremal correlators on flat space, in the analogous situation of 4D $\cN=2$ SCFTs, can be found in the appendix of \cite{Gerchkovitz:2016gxx}, the proof applies without any significant change to the 2D case as well.} We now go to a simpler scheme by doing a K\"ahler transformation:
	\beq
		Z_{S^2}^A( t, \ov t) \to \wt Z_{S^2}^A( t, \ov t) := \frac{Z_{S^2}^A( t, \ov t)}{X( t) \ov X(\ov  t)} = \frac{20}{3} e^{-2q \xi } \xi (-5\pi^2 + \xi^2)\,. \label{ZS2}
	\eeq
	Using $\pa_ t = -\frac{i}{2} \pa_\xi + \pi \pa_\tht$ and $\pa_{\ov  t} = \frac{i}{2} \pa_\xi + \pi \pa_\tht$ we can now compute flat space correlation functions using $\wt Z_{S^2}^A$ in \rf{key}. Next we find the ring relations.

\subsubsection{Relations and the ring}
	The correlation functions \rf{newscorr} are the matrix components from \rf{Matrix}:
	\beq
		M_{m,n} = \lan \mfr s_m(N) \ov{\mfr s}_n(S) \ran_{S^2} = \frac{\pa^m_ t \pa^n_{\ov  t} \wt Z_{S^2}^A}{\wt Z_{S^2}^A} = (-1)^m (i/2)^{m+n} \frac{\pa^{m+n}_\xi \wt Z_{S^2}^A}{\wt Z_{S^2}^A}\,, \label{S2excorrQ}
	\eeq
	where we could replace all the $t$ and $\ov t$-derivatives with $\xi$-derivatives because the partition function does not depend on the theta angle. Now the flat space correlators are given by:
	\beq
		\lan \si^m(0) \ov \si^n (\infty) \ran_{\bR^2} = \de_{m,n} 2^{2n} \frac{\det M_{(m)}}{\det M_{(m-1)}}\,. \label{R2excorrQ}
	\eeq
	We can get a recursion relation for $m>3$ and $n \ge 0$:
	\beq
		M_{m,n} = \frac{20}{3 \wt Z_{S^2}^A}  (-1)^m (i/2)^{m+n} \sum_{k=0}^3 (-2 q)^{m+n-k} e^{-2 q \xi} \pa^k_\xi (-5\pi^2 \xi + \xi^3) = i q M_{m-1,n}\,.
	\eeq
	This shows that for $m > 3$ the last $(m-2)$ rows of the matrix $M_{(m)}$ are multiples of each other and therefore $M_{(m)}$ has $(m-3)$ zero eigenvalues, i.e., $\det M_{(m)} \sim 0^{m-3}$. For the correlation functions the implication is:
	\beq
		\frac{0^{m-3}}{0^{m-4}} \sim \lan \si^m(0) \ov \si^m (\infty) \ran_{\bR^2} = 0 \,, \qquad \forall m > 3\,.
	\eeq
	Since the above correlation function is equivalent to an operator norm in a unitary theory, the operators $\si^m$ for $m>3$ themselves are identically zero. Thus we have fully determined the ring:
	\beq
		\mbox{Twisted chiral ring for the Quintic, } \cR^\tc = \bC[\si]/\lan\si^4\ran\,.
	\eeq
	This result was previously obtained in \cite{Morrison:1994fr,Closset:2015rna} using the topologically A-twisted version of the GLSM which doesn't require the supersymmetric localization and the counterterm analysis that we did. The upshot of going through the more elaborate method that we have presented is that this allows us to see the change in the ring structure as we move along the CFT moduli space; which we now discuss.

\subsubsection{Toda and $tt^*$-geometry of the bundle of BPS primaries}
	The nontrivial coupling dependence of the extremal correlators\footnote{the coupling dependence of the correlators \rf{R2excorrQ} derives from the coupling dependence of the partition function \rf{ZS2} via \rf{S2excorrQ}.} can be given a geometric interpretation \cite{deBoer:2008ss}, where we view the twisted chiral primaries as forming a holomorphic bundle with nontrivial connection over the conformal manifold parametrized by the exactly marginal coupling constants, the FI parameter in the present case. More generally, this picture applies to both chiral and twisted chiral rings in CFTs with arbitrary dimensional conformal manifolds. The CFT dynamics imposes the following curvature constraints (a.k.a. the $tt^*$ equations) on the geometry of these bundles:\footnote{The commutator of structure constants in \rf{tt*CC} looks like: $[C_\mu, \ov C_{\ov \nu}]_i^j = \tensor{C}{_{\mu i}^k} g_{k \ov l} \tensor{\ov C}{_{\ov \nu \ov m}^{\ov l}} g^{\ov m j} - g_{i \ov k} \tensor{\ov C}{_{\ov \nu \ov l}^{\ov k}} g^{\ov l m} \tensor{C}{_{\mu m}^j}$.}
	\begin{subequations}\begin{gather}
		[\nabla_\mu, \nabla_\nu]_i^j = [\ov \nabla_{\ov \mu}, \ov\nabla_{\ov \nu}]_i^j = 0\,, \\
		[\nabla_\mu, \ov\nabla_{\ov\nu}]_i^j = -[C_\mu, \ov C_{\ov \nu}]_i^j + g_{\mu\ov \nu} \de_i^j \lt(1 + \frac{R}{4c}\rt)\,. \label{tt*CC}
	\end{gather} \label{tt*} \end{subequations}
	Let us explain the notations: $\mu, \nu$ refer to tangential directions on the conformal manifold\footnote{i.e., they are indices for the exactly marginal couplings.} and $i, j$ refer to all BPS primaries. A bar over an index corresponds to an anti-BPS primary. In the second line, $R$ is the $U(1)_R$ charge of the bundle and $c$ is the central charge of the CFT. The matrices $C_\mu$, or more generally $C_i$, with indices expressed as $\tensor{C}{_{i j}^k}$ are the structure constants of our ring as defined in \rf{def:structC}, and finally, the metric $g_{\mu\ov\nu}$, or more generally $g_{i\ov j}$, is defined in terms of extremal correlators as in \rf{def:metric}.
	
	With a suitable choice for the basis of the BPS primaries over the conformal manifold, the $tt^*$ equation \rf{tt*CC} can be put into a more explicit form:\footnote{For details see \cite{Baggio:2014ioa} where the underlying theory was four dimensional but this portion of the computation applies to 2D as well.}
	\beq
		\frac{\pa}{\pa \ov \tau^{\ov \nu}} \lt( g^{\ov k j} \frac{\pa}{\pa \tau^\mu} g_{i \ov k} \rt) = [C_\mu, \ov C_{\ov \nu}]_i^j - g_{\mu\ov\nu} \de_i^j\,.
	\eeq
	Specializing to the case of a one dimensional conformal manifold, such as the Quintic (where the coordinate parametrizing the conformal manifold is $t$), and choosing basis of operators with diagonal structure constants and orthgonal metric $g_{i \ov j} = g_i \de_{ij}$,\footnote{Such as the basis $\{\si^i\}_{i=0}^3$ for the twisted chiral ring of the Quintic.} the above equation further simplifies to:
	\beq
		\pa_{\ov t} \pa_t \log g_k = \frac{g_{k+1}}{g_k} - \frac{g_k}{g_{k-1}} - g_1\,, \qquad k \in \{1, \cdots, \dim \cR-1\}, \quad g_{\dim\cR} = 0\,,
	\eeq
	where $\cR$ is the BPS ring of interest, for example, $\dim \cR_\tc = 4$ for the twisted chiral ring of the Quintic. The above equation is known as the Toda equation in the literature, which is usually written (after defining $q_k := \log g_k + Z_{S^2}$) in the more familiar form:
	\beq
		\pa_{\ov t} \pa_t q_k = e^{q_{k+1} - q_k} - e^{q_k - q_{k-1}}\,, \qquad k \in \{1, \cdots, \dim \cR-1\}, \quad q_{\dim \cR} = -\infty\,, \label{Toda}
	\eeq
	The condition $q_{\dim \cR} = -\infty$ signifies that the above equations are the equations of motion of a finite non-periodic Toda chain consisting of $\dim \cR$ sites located at $q_0, \cdots, q_{\dim \cR-1}$ where $q_k$ and is bound to $q_{k+1}$ by the potential $e^{q_{k+1}-q_k}$.
	
	The $tt^*$ equations \rf{tt*} are known to be integrable. In particular, the Toda equation \rf{Toda} can be solved explicitly given $g_1$ which we can compute using localization. Note that the Toda equation for the norms of the operators in an orthogonal basis with diagonal structure constant, in a one parameter theory, can be derived simply from the expression of these norms as a ratio of determinants (such as \rf{R2excorrQ}). An explicit derivation was presented in \cite{Gerchkovitz:2016gxx} in the context of 4d $\cN=2$ SCFT with $SU(2)$ gauge group, the proof remains unchanged for one parameter 2d BPS rings.

\subsection{Chiral Rings of the LG Minimal Models} \label{sec:LG}
	These are theories of chiral multiplets $X_i$, and the theories are characterized by a superpotential $W(X_i)$ and a K\"ahler potential $K(X_i,\overline X_i)$. If the superpotential has a quasi-homogenous singularity
	\beq
	W(\lambda^{m_i}X_i)=\lambda^2W(X_i)
	\eeq
	the Landau-Ginzburg model flows in the IR to a $(2,2)$ SCFT. The universality class of the SCFT is  insensitive to the choice of K\"ahler potential, which henceforth  we take to be canonical: $K(X, \ov X) = \frac{1}{2} \de^{ij} X_i \ov X_j$.

	\smallskip
	The equations of motion of the Landau-Ginzburg model gives: 
	\beq
	\partial_i W\propto \overline D^2 \overline X_i\,.
	\label{eom}
	\eeq	
	 Thus the bottom component of $\partial_i W$ is not only $\overline Q_B$-closed, but  also $\overline Q_B$-exact. Therefore  the chiral operator (the bottom component of)  $\partial_i W$ is represented by 0  in correlation functions with other chiral operators ($\partial_i W$ is not represented by 0 in an arbitrary correlator).
	 Therefore the chiral ring of the SCFT is the quotient
	\beq
	\cR^\ch = C[X_1,\ldots,X_n]/\lan \dd W \ran\,.	
	\label{chiralr}
	\eeq
	The chiral ring is spanned   by polynomials in the fields subject to the relations $dW=0$.   Unorbifolded Landau-Ginzburg models have a trivial twisted chiral ring. We will recover the result \rf{chiralr} from the sphere partition function in what follows.
	
	\smallskip 
		
	The $(2,2)$ unitary minimal models admit a Landau-Ginzburg description. The minimal model modular invariants pertain to  an ADE classification. This is mirrored by the following  ADE family of Landau-Ginzburg models:
	\beq\begin{aligned}
	A_k:&\qquad W=X^{k+1}\\\
	D_k:&\qquad W=X^{k-1}+XY^2\\
	E_6:&\qquad W=X^{3}+Y^4\\
	E_7:&\qquad W=X^{3}+XY^3\\
	E_8:&\qquad W=X^{3}+Y^5
	\end{aligned} \label{MM}\eeq
	The associated minimal models have the following properties:
	\begin{itemize}
	\item Central charge $c=3-{6\over h}$, where $h$ is the Coxeter number of the corresponding Lie group $G$.
	\item Dimension of the chiral ring, $\dim (\cR^\ch_G)=\text{rank}(G)$.
	\item The chiral operators $\cO_i \in {\cal R}_G^\ch$ have dimension $\Delta_i={d_i-2\over h}$, where $d_i$ is the order of the $i$-th Casimir of $G$.
	\end{itemize}
	Since $d_i\leq h$, all operators in the minimal models are relevant.

	\smallskip
	Extremal correlators of chiral and anti-chiral operators in a Landau-Ginzburg model on $S^2$ with a   superpotential $W(X_i)$ are given  by  \cite{Gomis:2012wy} (we set the radius of the sphere to 1)
	\beq
		\corrS{\cO_i(N) \overline \cO_{\ov j}(S)}=\int \prod_i dX_k d\overline X_{\ov k}\,
		\cO_i(X)\, \ov \cO_{\ov j}(\ov X)e^{-4\pi i(W(X) + \ov W(\ov X))\,.} 
	\eeq
	Recall from the general discussion of \S\ref{sec:exS2} that these are precisely the correlators that can be computed in background-B by localizing the path integral with respect to the supercharge $\cQ_B$.
	
\subsubsection{$A_{k+1}$ minimal model}
	We start first by analyzing the Landau-Ginzburg representation of the $A_{k+1}$ minimal model. The $su(2|1)_B$-invariant $S^2$ partition function for this  Landau-Ginzburg minimal model  is given by
	\begin{align}
		Z_{S^2}^{A_{k+1}} =&\; \int_{\bC} \dd X \dd\ov X\, e^{-4\pi iW(X) - 4\pi i \ov W(\ov X)} \nn\\
		=&\; \int_{\bR^2} \dd x\, \dd y\, e^{-4\pi i (x+iy)^{k+2} - 4\pi i (x-iy)^{k+2}} \nn\\
		=&\; \int_0^\infty \dd r \int_0^{2\pi} \dd\tht\, r e^{-8\pi i r^{k+2} \cos((k+2)\tht)}= \frac{\pi}{k+2} (4\pi)^{-\frac{2}{k+2}} {\Gamma\left({1\over k+2}\right)\over \Gamma\left({k+1\over k+2}\right)}\,.
		\label{minimalZ}
	\end{align}
	We would like to observe that this matches exactly  with the $su(2|1)_A$-invariant $S^2$ partition function of the same Landau-Ginzburg model. In this theory, the  $U(1)_V$-charge of $X$ is fixed by the superpotential to be ${2\over k+2}$, and it follows from the formulae in \cite{Benini:2012ui,Doroud:2012xw} that the $su(2|1)_A$-invariant partition function   is indeed \rf{minimalZ}. The physical interpretation of this equality of partition functions is mirror symmetry for the $k$-th minimal model MM$_k$, which exchanges
	\beq
	\hbox{MM}_k\Longleftrightarrow 	{\hbox{MM}_k\over Z_k}\,.
	\eeq
	 	
		\smallskip

		The ring relation $X^{k+1}=0$ can be   derived from the $S^2$ partition function. Indeed, using  the identity
	\beq
	\int dX d\overline X {d\over dX}e^{-4\pi i(X^{k+2}+\ov X^{k+2} +\bar t\, \ov X)}=0	\,,
	\eeq
	from which it follows that 
	\beq
	\vev{X^{k+1}(N) \ov X^\ell(S)}=0\qquad \forall \ell \,,
	\eeq
	implying that
	\beq
	X^{k+1}=0\,.
	\eeq
	This can be obtained from a differential equation. First we deform the supertotential by adding a source for the generator of the ring $W=X^{k+2}+ t X$. The two-sphere partition function then  obeys
	\beq
	\partial_t^{k+1} Z_{S^2}^{A_{k+1}}|_{t=0}=0\qquad \qquad\partial_{\bar t}^{k+1} Z_{S^2}^{A_{k+1}}|_{\bar t=0}=0\,.
	\eeq
	\smallskip
		
	The two-point functions of chiral operators are given by (we do the computation in \S\ref{sec:integral} using Riemann bilinear identity):
	\begin{subequations}\begin{align}
		M_{m,n}^{A_{k+1}} :=&\; \lt\lan X^m(N) \ov X^n(S) \rt\ran_{S^2} \nn\\
		=&\; \frac{1}{Z_{S^2}} \int_{\bC} \dd X \dd\ov X\, X^m \ov X^n e^{-4\pi iW(X) - 4\pi i \ov W(\ov X)} \label{MmnInt} \\
		=&\; \lt\{ \begin{array}{ccl} \frac{\pi(-i)^q}{(k+2)Z_{S^2}} (4\pi)^{-\frac{2(1+m)}{k+2}-q} \frac{\Ga\lt(\frac{m+1}{k+2}+q\rt)}{\Ga\lt(\frac{k-m+1}{k+2}\rt)} & \hspace{.5cm} & \mbox{if } q := \frac{n-m}{k+2} \in \bZ \\ 0 & \hspace{.5cm} & \mbox{otherwise} \end{array}  \rt. \label{Mmn}
	\end{align}\end{subequations}
	We see that any two operators with integer dimensions can mix.
	
	Recall that, $\Ga(-s)$ has poles at $s \in \bN_{\ge 0}$. Therefore \rf{Mmn} implies:
	\beq 
		M_{m,n}^{A_{k+1}} = 0 \quad \mbox{for} \quad \frac{k-m+1}{k+2} = -s\; \Rightarrow\; m = k+1+s(k+2) \quad \mbox{where} \quad s \in \bN_{\ge 0}\,.
	\eeq
	This in particular shows that, for $q=s=0$, $M_{k+1,k+1}^{A_{k+1}} = 0$. Now, let us denote by $M_{(n)}^{A_{k+1}}$ the matrix $M_{i,j}^{A_{k+1}}$ with $i,j = 0, \cdots, n$. Then $M_{(k+1)}^{A_{k+1}}$ is a diagonal matrix with exactly one zero row (the $(k+2)$'th row). This implies, using the Gram-Schmidt procedure (c.f. \rf{formula}),\footnote{In this particular case it is actually unnecessary to use the Gram-Schmidt procedure, because the operator $X^{k+1}$ has dimension $\frac{k+1}{k+2} < 1$ and the operators $\mathds 1, X, \cdots, X^{k+1}$ do not mix with each other (mixing can occur only at integer gaps in dimensions).}
	\beq
	\lt\lan X^{k+1}(0) \ov X^{k+1}(\infty) \rt\ran_{\bR^2} = 4^{\frac{k+1}{k+2}} \frac{\det M_{(k+1)}^{A_{k+1}}}{\det M_{(k)}^{A_{k+1}}} = 0
	\eeq
		By the Reeh-Schlieder theorem, we arrive at the ring relation
	\beq
	X^{k+1} = 0
	\eeq
	This implies that the chiral ring is given by:
	\beq
		\cR^\ch_{A_{k+1}} = \bC[X]/\lan X^{k+1} \ran\,.
	\eeq

		\smallskip
		
		It can also be explicitly checked that $\lt\lan X^m(0) \ov X^m(\infty)\rt\ran_{\bR^2} = 0$ for all $m > k$ implying that 
	\beq
	X^m=0 \qquad m>k\,.
	\eeq
	This can be shown as follows.  \rf{Mmn} allows to write a recursion relation between $M_{m,n}^{A_{k+1}}$ and $M_{m+k+2,n}^{A_{k+1}}$. Note that if we shift $m \to m+k+2$ then $m-n+q(k+2) = 0$ can be maintained by simultaneously shifting $q \to q-1$. Therefore,
		\begin{align}
			M_{m+k+2,n}^{A_{k+1}} =&\; \frac{\pi(-i)^{q-1}}{(k+2)Z_{S^2}} (4\pi)^{-\frac{2(1+m)}{k+2}-q+1} \frac{\Ga\lt(\frac{m+1}{k+2}+q\rt)}{\Ga\lt(\frac{k-m+1}{k+2}-1\rt)} \nn\\
			\Rightarrow \frac{M_{m+k+2,n}^{A_{k+1}}}{M_{m,n}^{A_{k+1}}} =&\; -4\pi i \lt(\frac{m+1}{k+2}\rt)
		\end{align}
		This implies that for any $m > k+1$ the $(m+1)$'th row of $M_{(m)}^{A_{k+1}}$ is a multiple of the $(m-k-1)$'th row of $M_{(m)}^{A_{k+1}}$. Therefore, for $m>k$, the number of 0 eigenvalues of $M_{(m)}^{A_{k+1}}$ is $m-k$ (note that $M_{(k+1)}^{A_{k+1}}$ already has one zero eigenvalue). Hence:
		\beq
			\lt\lan X^m(0) \ov X^m(\infty)\rt\ran_{\bR^2} = 2^{\frac{2m}{k+2}} \frac{\det M_{(m)}^{A_{k+1}}}{\det M_{(m-1)}^{A_{k+1}}} \sim \frac{0^{m-k}}{0^{m-k-1}} = 0\,, \qquad \mbox{for} \quad m > k\,.
		\eeq
	\smallskip
	
	Finally, we can make contact with previously known results about the OPE structure of these chiral primaries obtained from CFT methods \cite{Cecotti:1989jc, Mussardo:1988av}. First, let us normalize the chiral ring operators by defining $\wh{X^n} := X^n/\lVert X^n \rVert$ where $\lVert X^n \rVert^2 = \corrR{X^m(0) \ov X^m(\infty)}$. Then using the relation $X^m X^n = X^{m+n}$ we can compute the OPE coefficients for the normalized operators:
	\beq
		\wh{ X^m} \wh{ X^n} = \cF_{m,n}^{A_{k+1}} \wh{ X^{m+n}}\,, \qquad \cF_{m,n}^{A_{k+1}} = \frac{\lVert X^{m+n}\rVert}{\lVert X^m \rVert \lVert X^n \rVert}\,. \label{OPEA}
	\eeq
	These OPE coefficients depend only on the dimensions of the operators and the central charge $c$ \cite{Cecotti:1989jc, Mussardo:1988av}. For the ADE models, the central charge $c = 3 - \frac{6}{h}$ where $h$ is the Coxeter number of the corresponding Lie group. For example, the Coxeter number of $A_{k+1}$ is $(k+2)$, so the central charge of the $A_{k+1}$ model is:
	\beq
		c_{A_{k+1}} = \frac{3k}{k+2}\,.
	\eeq
	The central charge dependence of the OPE coefficients is implicit in  \rf{OPEA} and can be seen through the dependence of the correlation functions on $k$ (see \rf{Mmn}). These OPE coefficients can be computed using \rf{OPEA}, \rf{Mmn}, and \rf{formula}, the result is:
	\beq
		\cF_{m,n}^{A_{k+1}} = \sqrt{\frac{\Ga\lt(\frac{1}{k+2}\rt) \Ga\lt(\frac{k-m+1}{k+2}\rt) \Ga\lt(\frac{k-n+1}{k+2}\rt) \Ga\lt(\frac{m+n+1}{k+2}\rt)}{\Ga\lt(\frac{k+1}{k+2}\rt) \Ga\lt(\frac{m+1}{k+2}\rt) \Ga\lt(\frac{n+1}{k+2}\rt) \Ga\lt(\frac{k-m-n+1}{k+2}\rt)}}\,. \label{AOPEco}
	\eeq
	These coefficients can also be read off from the results of \cite{Cecotti:1989jc, Mussardo:1988av} and they match with our expression.
	
	

\subsubsection{$D_{k+1}$ minimal model}
	The same analysis can be carried out for the $D_{k+1}$ minimal model in a completely analogous way. The superpotential for the LG theory describing this minimal model is:
	\beq
		W(X,Y) = X^k + XY^2\,,
	\eeq
	with two generators $X$ and $Y$ corresponding to conformal primaries of dimensions $\frac{1}{k}$ and $\frac{k-1}{2k}$ respectively. With the canonical kinetic Lagrangian $\int \dd^4\tht (X\ov X + Y \ov Y)$ the equations of motion tells us:
	\beq
		\ov D^2 \ov X \propto k X^{k-1} + Y^2\,, \qquad \ov D^2 \ov Y \propto X Y\,.
	\eeq
	Therefore, the chiral ring can be described as:
	\beq
		\cR^\ch_{D_{k+1}} = \bC[X,Y]/\lan k X^{k-1} + Y^2, XY \ran\,. \label{chringD}
	\eeq
	We will derive these ring relations by computing correlation functions of the generators and showing that the operators $k X^{k-1} + Y^2$ and $XY$ are zero in the chiral ring. Note that according to the relations, a minimal (dimension-wise) basis for the ring is given by:
	\beq
		\mathds 1, X, \cdots, X^{k-1}, \quad \mbox{and} \quad Y\,. \label{basisD}
	\eeq
	Here $X^{k-1}$ has the highest dimension, $\frac{k-1}{k} < 1$. Based on our supergravity analysis we can expect that operators can only mix at integer gaps in dimensions,\footnote{As we found explicitly for the $A_{k+1}$ model following \rf{Mmn}.} in fact we will establish this by explicit computation. This implies that there is no mixing among the operators in the minimal basis, simplifying our computations.
	
	When put on a two-sphere, a general extremal correlation function in this LG model is given by:
	\begin{align}
		M_{m,n,p,q}^{D_{k+1}} :=&\; \corrS{X^m(N) Y^n(N) \ov X^p(S) \ov Y^q(S)} \nn\\
		=&\; \frac{1}{Z_{S^2}^{D_{k+1}}} \int_{\bC^2} \dd X \dd \ov X \dd Y \dd \ov Y\, X^m Y^n \ov X^p \ov Y^q e^{-4\pi i \lt(X^k+XY^2 + \ov X^k + \ov X \ov Y^2\rt)} \label{corrD}
	\end{align}
	We can carry out the $Y, \ov Y$ integrals first:
	\begin{align}
		\int_\bC \dd Y \dd \ov Y\, Y^n \ov Y^q e^{-4\pi i(XY^2 + \ov X \ov Y^2)} =&\; X^{-\frac{n+1}{2}} \ov X^{-\frac{q+1}{2}} \int_\bC \dd Y \dd \ov Y\, Y^n \ov Y^q e^{-4\pi i(Y^2 + \ov Y^2)}\,, \nn\\
		=&\; X^{-\frac{n+1}{2}} \ov X^{-\frac{q+1}{2}} Z_{S^2}^{A_1} M_{n,q}^{A_1}\,. \qquad \mbox{[c.f. \rf{MmnInt}]}
	\end{align}
	Then we can perform the integrals over $X, \ov X$, which leads to an expression for the $D$-type extremal correlators in terms of the $A$-type extremal correlators:
	\beq
		M_{m,n,p,q}^{D_{k+1}} = \frac{Z_{S^2}^{A_1} Z_{S^2}^{A_{k-1}}}{Z_{S^2}^{D_{k+1}}} M_{n,q}^{A_1} M_{m-\frac{n+1}{2}, p-\frac{q+1}{2}}^{A_{k-1}}\,. \label{corrDx1}
	\eeq
	Let us define some symbols:
	\beq
		\wt m := m - \frac{n+1}{2}\,, \quad \wt p := p - \frac{q+1}{2}\,, \quad \up := \frac{q-n}{2}\,, \quad \chi := \frac{\wt p-\wt m}{k}\,,
	\eeq
	then we can write \rf{corrDx1} more explicitly as (c.f. \rf{Mmn}):
	\beq
		M_{m,n,p,q}^{D_{k+1}} = \lt\{\begin{array}{ll} \frac{1}{Z_{S^2}^{D_{k+1}}} \frac{(-i)^{\up+\chi}}{32k} (4\pi)^{1-n-\up-\chi - \frac{2}{k}(\wt m +1)}
			\frac{\Ga\lt(\frac{1+n}{2}+\up\rt) \Ga\lt(\frac{1+ \wt m}{k} + \chi\rt)}{\Ga\lt(\frac{1-n}{2}\rt) \Ga\lt(\frac{k-\wt m -1}{k}\rt)} & \hspace{.5cm} \up, \chi \in \bZ \\
			0 & \hspace{.5cm} \mbox{otherwise}
			\end{array}\rt. \label{corrDx}
	\eeq
	Noting the difference in dimension:
	\beq
		\varDelta(X^mY^n) - \varDelta(X^pY^q) = \up + \chi\,,
	\eeq
	we observe that mixing between different operators can occur on the sphere only if their dimensions differ by an integer amount.  Since all the operators in the minimal basis \rf{basisD} have dimensions less than $1$, there is no mixing among them. Therefore, their correlation functions on the sphere of unit radius, namely the ones given by \rf{corrDx}, are simply proportional to the corresponding flat space correlators.
	
	In order to check the ring relations, we define the following two operators:
	\beq
		\cO_1 := k X^{k-1} + Y^2, \quad \mbox{and} \quad \cO_2 := XY\,.
	\eeq
	
	Using the explicit form of the correlation functions \rf{corrDx} we can easily check that these two operators have zero norms:
	\beq\begin{aligned}
		\lVert \cO_1 \rVert^2 =&\; \corrR{\cO_1(0) \ov\cO_1(\infty)} = 4^{\frac{k - 1}{k}} \corrS{\cO_1(N) \ov\cO_1(S)} \\
		=&\; 4^{\frac{k - 1}{k}}  \lt(k^2 M_{k-1,0,k-1,0}^{D_{k+1}} + k M_{k-1,0,0,2}^{D_{k+1}} + k M_{0,2,k-1,0}^{D_{k+1}} + M_{0,2,0,2}^{D_{k+1}} \rt) = 0\,,
	\end{aligned}\eeq
	and similarly,
	\beq
		\lVert \cO_2 \rVert^2 = \corrR{\cO_2(0) \ov\cO_2(\infty)} = 2^{\frac{k+1}{k}} \corrS{\cO_2(N) \ov\cO_2(S)} = 2^{\frac{k+1}{k}} M_{1,1,1,1}^{D_{k+1}} = 0\,.
	\eeq
	By the   Reeh-Schlieder theorem, we arrive at the ring relations:
	\beq
		k X^{k-1} + Y^2 = XY = 0\,,
	\eeq
	giving us the chiral ring \rf{chringD}.
	\smallskip
	
	Once we normalize the chiral primaries, $\wh{X^mY^n} := \frac{X^mY^n}{\lVert X^mY^n \rVert}$, we can compute the OPE coefficients for the product of two arbitrary chiral primaries in terms of the extremal correlators (as we did for the $A$ series in \rf{OPEA} and \rf{AOPEco}).
	
	Let us make a few remarks about the computation. We define the OPE coefficients of the normalized operators by the following equation:
	\beq
		\wh{X^mY^n} \wh{X^pY^q} = \cF_{(m,n),(p,q)}^{D_{k_2}} \wh{X^{m+n} Y^{p+q}}\,, \qquad \cF_{(m,n),(p,q)}^{D_{k_2}} = \frac{\lt\lVert X^{m+n} Y^{p+q} \rt\rVert}{\lt\lVert X^m Y^n \rt\rVert \lt\lVert X^p Y^q \rt\rVert}\,. \label{OPEB}
	\eeq
	As we mentioned after \rf{OPEA}, these coefficients depend only the dimensions of the operators and the central charge, in particular, they should be computable without making a choice a modular invariant. Therefore we should expect the OPE coefficients defined in \rf{OPEA} and the ones in \rf{OPEB} to be the same:
	\beq
		\cF_{m,n}^{A_{k+1}} = \cF_{(p,q),(r,s)}^{D_{k'+1}}\,, \label{coA=coD}
	\eeq
	whenever the central charges of the two theories and the dimensions of the involved operators coincide, i.e.:\footnote{The Coxeter numbers of $A_{k+1}$ and $D_{k+1}$ are $(k+2)$ and $2k$ respectively. Two ADE models have the same central charge as long as the corresponding Lie groups have the same Coxeter number.}
	\begin{subequations}\begin{align}
		c_{A_{k+1}} = c_{D_{k'+1}} \quad \Rightarrow&\; \quad k' = \frac{k+2}{2} \,, \\
		\De_{A_{k+1}}(X^m) = \De_{D_{k'+1}}(X^p Y^q) \quad \Rightarrow&\; \quad m = \frac{1}{2}(4p+kq) \,, \\
		\De_{A_{k+1}}(X^n) = \De_{D_{k'+1}} (X^r Y^s)\,, \quad \Rightarrow&\; \quad n = \frac{1}{2}(4r+ks) \,.
	\end{align}\label{matchCond}\end{subequations}
	Explicit computation shows that the equality \rf{coA=coD} indeed holds given that the above conditions are satisfied.

\subsubsection{Exceptional minimal models}
	Our discussion so far for the $A$ and the $D$-series of minimal models readily extends to the $E_6$, $E_7$ and $E_8$ models. Like the $D$-series, the chiral rings of these models have two generators. The $E_6$ and the $E_8$ minimal model superpotentials, namely $X^3 + Y^4$ and $X^3 + Y^5$ \rf{MM}, are decoupled sums of two polynomials in these two generators and therefore the chiral rings of these two models are simply Cartesian products of two $A$-type chiral rings:
	\beq
		\cR^\ch_{E_6} = \cR^\ch_{A_2} \times \cR^\ch_{A_3}\,, \qquad \cR^\ch_{E_8} = \cR^\ch_{A_2} \times \cR^\ch_{A_4}\,.
	\eeq
	
	The $E_7$ model involves nontrivial coupling between the two generators analogous to the $D$-series. Recall that the $E_7$ superpotential is $X^3 + XY^3$ and therefore the chiral ring is:
	\beq
		\cR^c_{E_7} = \bC[X,Y]/\lan 3X^2 + Y^3, XY^2 \ran\,. \label{e7ring}
	\eeq
	There are two chiral primaries $X$ and $Y$ of dimensions $\frac{1}{3}$ and $\frac{2}{9}$ respectively. A basis for the chiral ring is given by $\{\mathds 1, Y, X, Y^2, XY, Y^3, X^2 Y\}$ in increasing order of dimension. The highest dimensional chiral primary $X^2Y$ has dimension $\frac{8}{9} < 1$ and therefore there is no mixing among these basis operators when we put the corresponding LG theory on a sphere. Thus once again we have proportionality between extremal correlators on a sphere of unit radius and extremal correlators on $\bR^2$. Just as we did in \rf{corrDx1}, we can write down the extremal correlators in this model in terms of $A$-series extremal correlators:
	\beq
		M_{m,n,p,q}^{E_7} := \corrS{X^m(N) Y^n(N) \ov X^p(S) \ov Y^q(S)} = \frac{Z_{S^2}^{A_2} Z_{S^2}^{A_2}}{Z_{S^2}^{E_7}} M_{n,q}^{A_2} M_{m - \frac{n+1}{3}, p - \frac{q+1}{3}}^{A_2}\,.
	\eeq
	Using the explicit form of the $A$-type extremal correlators \rf{Mmn}, we can verify that in the $E_7$ model the operators $\cO_1 := 3X^2 + Y^3,\, \cO_2 := XY^2$ have zero norms:
	\beq\begin{aligned}
		\lVert \cO_1 \rVert^2 =&\; 2^{\frac{4}{3}} \lt( 9 M_{2,0,2,0}^{E_7} + 3 M_{2,0,0,3}^{E_7} + 3 M_{0,3,2,0}^{E_7} + M_{0,3,0,3}^{E_7} \rt) = 0\,, \\
		\lVert \cO_2 \rVert^2 =&\; 2^{\frac{14}{9}} M_{1,2,1,2}^{E_7} = 0\,.
	\end{aligned}\eeq
	This establishes the ring relations in \rf{e7ring} by identifying $\cO_1$ and $\cO_2$ with the null operator. 

\section{Epilogue}
	In this paper we have approached the computation of 2D BPS ring structure constants from the conceptually straightforward way of computing the relevant correlation functions. The main tool at our disposal was supersymmetric localization allowing exact computation of these correlation functions on the sphere. The main obstacle in using these results to compute the ring structure is the presence of conformal anomalies on the sphere leading to operator mixing. We have explored the roots of this mixing in the supergravitational descriptions of the sphere, and we have outlined a way of computing the flat space correlators from the sphere partition function in the presence of such mixing. We have demonstrated our method in some familiar theories of interest and reproduced known results about the structure constants that were previously obtained via CFT methods and we have also verified all the ring relations. This work builds on previous works in 4D and extends the applicability and usefulness of the already exceptionally useful results from localization while providing a simple elementary perspective on the BPS algebraic structures of 2D $\cN=(2,2)$ theories.

\section*{Acknowledgment}
	We thank Jaume Gomis for motivating and providing core insights behind this project, as well as for many essential discussions on the subject.

	\appendix

\section{The $\cN=(2,2)$ superconformal Algebra} \label{sec:algebra}

	We use complex coordinates on $\bR^2$:
	\beq
		z = x + iy\,, \quad \ov z = x-iy\,.
	\eeq
	The $\cN = (2,2)$ superconformal algebra\footnote{We are only concerned with algebra that is globally defined on $\bR^2$, i.e., the globally defined subalgebra of the super Virasoro algebra.} $su(2|2)$ contains eight supercharges:
	\beq
		Q_+\,, Q_-\,, \ov Q_+\,, \ov Q_-\,, \quad \mbox{and,} \quad S_+\,, S_-\,, \ov S_+\,, \ov S_-\,, \label{QS}
	\eeq
	and the following bosonic generators:
	\beq\begin{aligned}
		\mbox{Rotation in } \bR^2,\; u(1)_L :\quad & 2(\ov L_0 - L_0) =: 2J_L \\
		\mbox{Dilatation} :\quad & L_0 + \ov L_0  =: \De \\
		\mbox{Translations} :\quad & L_{-1}, \ov L_{-1} \\
		\mbox{Special conformal transformations} :\quad & L_1, \ov L_1 \\
		\mbox{Vector R-symmetry, } u(1)_V :\quad & J_V \\
		\mbox{Axial R-symmetry, } u(1)_A :\quad & J_A
	\end{aligned}\label{gens}\eeq
	where $L_s := -z^{s+1} \pa_z$ and $\ov L_s := - \ov z^{s+1} \pa_{\ov z}$ gnerate the conformal algebra $\mathrm{conf}(\bR^2) = so(3,1)$:
	\beq
		[L_r, L_s] = (r-s) L_{r+s}\,, \quad [\ov L_r, \ov L_s] = (r-s) \ov L_{r+s}\,, \qquad r, s \in \{-1,0,1\}
	\eeq
	The nonzero anti-commutation relations of the supercharges \rf{QS} are:
	\begin{subequations}\begin{align}
		\{Q_+, \ov Q_+\} = 2L_{-1}\,, \qquad& \{Q_-, \ov Q_-\} = 2\ov L_{-1}\,, \label{QQL-1} \\
		\{S_+, \ov S_+\} = 2L_1\,, \qquad& \{S_-, \ov S_-\} = 2 \ov L_1\,, \\
		\{Q_+, \ov S_+\} = 2L_0 + \frac{1}{2}(J_V + J_A)\,, \qquad& \{Q_-, \ov S_-\} = 2 \ov L_0 + \frac{1}{2}(J_V - J_A)\,, \\
		\{\ov Q_+, S_+\} = 2L_0 - \frac{1}{2}(J_V + J_A)\,, \qquad& \{\ov Q_-, S_-\} = 2 \ov L_0 - \frac{1}{2}(J_V - J_A)\,.
	\end{align}\label{QQ}\end{subequations}
	The commutators of the supercharges with the $u(1)$'s and the dilatation are conveniently expressed by specifying the charges of the supercharges under the respective generators:
	\beq
	\begin{array}{c|c|c|c|c}
		& Q_\pm & \ov Q_\pm & S_\pm & \ov S_\pm \\
		\hline
		2J_L & \mp & \mp & \pm & \pm \\
		J_V & - & + & - & + \\
		J_A & \mp & \pm & \mp & \pm \\
		2\De & + & + & - & -
	\end{array}
	\eeq
	The rest of the nonzero commutators of $su(2|2)$ are:
	\beq
		\begin{array}{cccc}
			\,[L_1, Q_+] = S_+\,, \quad & [L_1, \ov Q_+] = \ov S_+\,, \quad & [L_{-1}, S_+] = -Q_+\,, \quad & [L_{-1}, \ov S_+] = -\ov Q_+\,,  \\
			\,[\ov L_1, Q_-] = S_-\,, \quad & [\ov L_1, \ov Q_-] = \ov S_-\,, \quad & [\ov L_{-1}, S_-] = -Q_-\,, \quad & [\ov L_{-1}, \ov S_-] = -\ov Q_-\,.
		\end{array}
	\eeq

\section{Supersymmetry on the sphere} \label{app:S2susy}
	
	A theory with $\cN=(2,2)$ superconformal symmetry, namely the symmetry algebra $su(2|2)$, can be put on the two-sphere by a Weyl transformation, classically preserving the full superconformal symmetry. Though UV regularization will break the $su(2|2)$ symmetry to an $su(2|1)$ subalgebra. On the other hand, a nonconformal theory, such as a guage theory, can preserve even classically only an $su(2|1)$ subalgebra of the full $su(2|2)$ superconformal algebra.
	\smallskip
	
	$su(2|1)$ subalgebras of $su(2|2)$ consist of the isometries of the two-sphere, supercharges that generate these isometries and a $u(1)$ subalgebra of the $u(1)_V \times u(1)_A$ R-symmetry algebra of $su(2|2)$.\footnote{If the theory is to flow to a nontrivial CFT in the IR the theory must preserve both of the $U(1)$ R-symmetries.} There are two non-equivalent\footnote{Not related by any inner automorphism of $su(2|2)$.} $su(2|1)$ subalgebras of $su(2|2)$, one of them contains the vector R-symmetry $u(1)_V$ and the other one contains the axial R-symmetry $u(1)_A$. They are referred to as $su(2|1)_A$ and $su(2|1)_B$ respectively.\footnote{Note the slightly unfortunate notation that $u(1)_A$ is contained in $su(2|1)_B$ and not in $su(2|1)_A$.}
	\smallskip
	
	The two supersymmetric sphere backgrounds can be derived as two different supergravity backgrounds preserving four supercharges. We will refer to the $su(2|1)_A$ and $su(2|1)_B$ preserving sphere backgrounds as ``background-A" and ``background-B" respectively. The conformal Killing spinor equations are:
	\beq
		\nabla_m \ep(x) = \eta(x)\,, \qquad \nabla_m \wt \ep(x) = \wt\eta(x)\,, \label{CKS}
	\eeq
	where $\ep$, $\wt\ep$, $\eta$ and $\wt\eta$, parametrize the $\ov Q$, $Q$, $\ov S$, and $S$ transformations. The two sphere backgrounds are defined by imposing constraints on the $S$-supersymmetris as we discuss in the following.

\subsubsection{Background-A}
	The Killing spinor equation of the supersymmetric $S^2$ background preserving the vector R-symmetry is found by imposing on $\eta$ and $\wt\eta$ the follwoing constraints \cite{Gerchkovitz:2014gta,Benini:2012ui,Closset:2014pda,Bae:2015eoa}:
	\beq
		\eta = \frac{i}{2r} \ep\,, \qquad \wt\eta = \frac{i}{2r} \wt\ep\,.
	\eeq
	So that the Killing spinor equations end up being:
	\beq
		\nabla_m \ep(x) = \frac{i}{2r} \ga_m \ep(x)\, \qquad \nabla_m \wt\ep(x) = \frac{i}{2r} \ga_m \wt\ep(x)\,, \label{KSEA}
	\eeq
	where $r$ is the radius of the sphere, and the covariant derivative $\nabla_m$ does not contain any background field other than the spin connection. The Killing spinor equations \rf{KSEA} have a (complex) four dimensional space of solutions that can be written as:\footnote{We are using $\ga$ and $\Ga$ to refer to the curved space and flat space gamma matrices respectively. In stereographic coordinate the metric on the sphere of radius $r$ is $\lt(1 + \frac{x^2}{4r^2} \rt)^{-2} \mathrm{diag}(1, 1)$.}
	\begin{subequations}\begin{align}
		\ep_{\chi_0, \wt\chi_0}^A(x) =&\; \frac{1}{\sqrt{1 + \frac{x^2}{4r^2}}} \lt(\mathds{1} + \frac{i}{2r} x^m \Ga_m \rt) \chi_0\,, \\
		\wt\ep_{\chi_0, \wt\chi_0}^A(x) =&\; \frac{1}{\sqrt{1 + \frac{x^2}{4r^2}}} \lt(\mathds{1} + \frac{i}{2r} x^m \Ga_m \rt) \wt\chi_0\,. \label{solKSEAb}
	\end{align}\label{solKSEA}\end{subequations}
	Here $\chi_0$ and $\wt\chi_0$ are two constant Dirac spinors parametrizing the space of solutions.

\subsubsection{Background-B}
	Analogously, the axial R-symmetry preserving background is defined by imposing in the conformal Killing spinor equation \rf{CKS} \cite{Gerchkovitz:2014gta}:
	\beq
		\eta = \frac{i}{2r} \wt\ep\,, \qquad \wt\eta = \frac{i}{2r} \ep\,,
	\eeq
	so that we have the following Killing spinor equations:
	\beq
		\nabla_m \ep(x) = \frac{i}{2r} \ga_m \wt\ep(x)\, \qquad \nabla_m \wt\ep(x) = \frac{i}{2r} \ga_m \ep(x)\,. \label{KSEB}
	\eeq
	These can be solved by defining:
	\beq
		\vep := \ep_+ + \wt\ep_-\,, \qquad \wt\vep := \ep_- + \wt\ep_+ \label{def:vep}
	\eeq
	which satisfy the already solved equations \rf{KSEA}:
	\beq
		\nabla_m \vep = \frac{i}{2r} \ga_m \vep\, \qquad \nabla_m \wt\vep = \frac{i}{2r} \ga_m \wt\vep\,.
	\eeq
	Thus we find that the solutions to \rf{KSEB} are given by:
	\begin{subequations}\begin{align}
		\ep_{\chi_0, \wt\chi_0}^B(x) =&\; \frac{1}{\sqrt{1 + \frac{x^2}{4r^2}}} \lt(\chi_{0+} + \wt\chi_{0-} + \frac{i}{2r} x^m \Ga_m (\chi_{0-} + \wt\chi_{0+}) \rt)\,, \\
		\wt\ep_{\chi_0, \wt\chi_0}^B(x) =&\; \frac{1}{\sqrt{1 + \frac{x^2}{4r^2}}} \lt(\chi_{0-} + \wt\chi_{0+} + \frac{i}{2r} x^m \Ga_m (\chi_{0+} + \wt\chi_{0-}) \rt)\,,
	\end{align} \label{solKSEB}\end{subequations}
	parametrized by two constant Dirac spinors $\chi_0$ and $\wt\chi_0$.

\section{Ward identity} \label{app:Ward}
	Let us present the proof of the Ward identity from \S\ref{sec:Ward} for twisted chiral multiplets in background-A. 
\subsubsection{Supersymmetric deformations of the action}
	For a twisted chiral primary $Y$ of Weyl weight $w$, the $su(2|1)_A$ variations of the twisted chiral multiplet $\Psi = (Y, \ze, G)$, generated by the Killing spinors $\ep$ and $\wt\ep$, are \cite{Gerchkovitz:2014gta}:
	\begin{subequations}
	\begin{align}
		\de Y =&\; \wt\ep_+ \ze_- - \ep_- \ze_+\,, \\
		\de \ze_+ =&\; -i \slashed \pa Y \wt\ep_- + \lt( G + \frac{w}{r} Y \rt) \wt\ep_+ \,, \label{su21Atcb} \\
		\de \ze_- =&\; i \slashed \pa Y \ep_+ - \lt( G + \frac{w}{r} Y \rt) \ep_- \,, \\
		\de G =&\; -i \wt\ep_- \slashed \nabla \ze_- + i \ep_+ \slashed \nabla \ze_+ + \frac{w}{r} \lt( \ze_+\ep_- - \ze_- \wt\ep_+\rt) \,.
	\end{align}\label{su21Atc}
	\end{subequations}
	The twisted F-term action for $\Psi$ on the sphere invariant under these variations is given by \cite{Gomis:2012wy, Closset:2014pda}:
	\beq
		I_{w}^\mathrm{tc-F}(\Psi) := \int_{S^2} \dd^2x \sqrt{g(x)}\, \cG(\Psi) \,, \qquad \cG(\Psi) = G + \frac{w-1}{r}Y. \label{tcF}
	\eeq
	The subscript on $I$ refers to the Weyl weight of $Y$, the Weyl weight of the integral $I_w^\mathrm{tc-F}$ is $w-1$, therefore a deformation of an action $S$ by this term can be introduced by simply introducing a coupling $\tau$ of Weyl weight $(1-w)$:
	\beq
		S \to S - \frac{i\tau}{4\pi} I_{w}^\mathrm{tc-F}(\Psi)\,. \label{tcFdef}
	\eeq
	
	We want to show that the integrated operator $I_{w}^\mathrm{tc-F}(\Psi)$ localizes to a point inside an extremal correlator. To proceed, let us pick a particular supercharge $Q_A \in su(2|1)_A$ by restricting $\wt\chi_0$ in \rf{solKSEA} to be chiral:
	\beq
		\wt\chi_{0+} = 0\,.
	\eeq
	Then the Killing spinor \rf{solKSEAb}, which we write simply as $\wt\ep$, becomes:
	\beq
		\wt\ep_+ = \frac{1}{\sqrt{1 + \frac{x^2}{4r^2}}} \frac{i}{2r} x^m \Ga_m \wt\chi_{0-}\,, \qquad
		\wt\ep_- = \frac{1}{\sqrt{1 + \frac{x^2}{4r^2}}} \wt\chi_{0-}\,. \label{Q}
	\eeq
	Solving \rf{su21Atcb} for $G$ and substituting it in the expression for $\cG$ in \rf{tcF}, we find:
	\beq
		\cG(\Psi) = \de \lt( \frac{\wt\ep_+^\dagger \ze_+}{\lVert \wt\ep_+ \rVert^2} \rt) + \frac{i}{\lVert \wt\ep_+ \rVert^2} \wt\ep_+^\dagger \slashed\nabla(Y \wt\ep_-)\,.
	\eeq
	were $\lVert \wt\ep_+ \rVert^2 := \wt\ep_+^\dagger \wt \ep_+$. Using complex coordinates $z = x^1 + i x^2$ and $\ov z = x^1 - i x^2$ (so that $\wt\ep_+$ satisfies the simple equations $\nabla_z \wt\ep_+ = 0$ and $\nabla_z \frac{\wt\ep_+^\dagger}{\lVert \wt\ep_+ \rVert^2} = 0$) we can turn the second term into a total derivative:
	\beq
		\cG(\Psi) = \de \lt( \frac{\wt\ep_+^\dagger \ze_+}{\lVert \wt\ep_+ \rVert^2} \rt) + 2i \nabla_z \lt( \frac{1 + \frac{z \ov z}{4r^2}}{\lVert \wt\ep_+ \rVert^2}\, \wt\ep_+^\dagger \Ga_1 \wt\ep_- Y \rt)\,. \label{G}
	\eeq
	The key point here is that the norm of $\wt\ep_+$ vanishes at $z = \ov z = 0$ (see \rf{Q}) and therefore the space-time integral of the derivative localizes at the origin, which we call the North pole $N$. When inserted in a correlator with $Q_A$-closed operators, such as an extremal correlator, the $Q_A$-exact term in \rf{G} can be ignored. In \cite{Gerchkovitz:2014gta} the integral of the derivative was computed to be $-4\pi r Y(N)$. Therefore we reach the conclusion that inside an extremal correlator:
	\beq
		\corrS{\int_{S^2} \dd^2x \sqrt{g(x)}\, \cG(\Psi) \cdots} = -4\pi r\corrS{Y(N) \cdots}\,.
	\eeq
	This is the equation \rf{wardtc}. The equation \rf{wardtac} can be proven by starting from the $su(2|1)_A$-variation of a twisted anti-chiral multiplet. The analogous equations for the chiral and the anti-chiral multiplets in background-B are proven similarly.

	\section{Contour integrals} \label{sec:integral}
		On the two-sphere, the extremal correlation functions of chiral operators in a LG model of type $A_{k+1}$ involve the following integrals (c.f. \S\ref{sec:LG}):
		\beq
			\int_\bC \dd X \dd\ov X\, X^m \ov X^n e^{-4\pi i X^{k+2} - 4\pi i \ov X^{k+2}} = (4\pi)^{-\frac{2(m+1)}{k+2} - q} (-i)^q \int_\bC \dd X \dd\ov X\, X^m \ov X^n e^{X^{k+2} - \ov X^{k+2}}\,, \label{XmXn1}
		\eeq
		where $q = \frac{n-m}{k+2}$. After defining:
		\beq
			\om_m := X^m e^{X^{k+2}}\,\dd X\,, \qquad \wt\om_m := \ov X^m e^{-\ov X^{k+2}}\,\dd \ov X\,,
		\eeq
		we can write:\footnote{The factor of $i/2$ is there because by $\dd X \dd \ov X$ we mean $\dd x \dd y = \dd x \wedge \dd y$ where $X = x+iy$ and $\ov X = x-iy$, whereas $\dd X \wedge \dd \ov X = (\dd x + i \dd y) \wedge (\dd x - i \dd y) = -2i \dd x \wedge \dd y$.}
		\beq
			\int_\bC \dd X \dd\ov X\, X^m \ov X^n e^{X^{k+2} - \ov X^{k+2}} = \frac{i}{2}\int_\bC \om_m \wedge \wt\om_n\,. \label{XmXn2}
		\eeq
		We will evaluate this integral by writing it as a sum of integrals of $\om_m$ and $\wt\om_n$ over one-cycles that we will define momentarily. This procedure will use a generalization of the Riemann bilinear identity as explained in Appendix C of \cite{Gaiotto:2013rk}.\footnote{The general idea behind evaluating certain integrals by decomposing them over some cycles comes from Picard–Lefschetz theory, which has been used in the past to compute integrals similar to ours\cite{Cecotti:1992rm, Witten:2010cx, Hori:740255}. Also note that the integration \rf{XmXn1} seems to be computable using integral identity involving Bessel function of the first kind, but we were unable to confirm that all convergence conditions relevant for the integral identity are satisfied in the present case. Still, we note that a straightforward application of the Bessel function identity produces exactly the same result as the one given by the Riemann bilinear identity.}
		
		We denote by $\tht_\uparrow$ a ray originating from the origin of the complex plane and going off to infinity at an angle $\tht$ with the $x$-axis. And by $\cC_{\tht_1,\tht_2}$ we will refer to a curve that originates at infinity, comes near the origin and then goes off to infinity again in a way such that it is wedged between the rays $\tht_{1\uparrow}$ and $\tht_{2\uparrow}$ and approaches these two rays asymptotically. Such curves will be thought of as noncompact cycles in the complex plane. Our convention is such that $\tht_{2\uparrow}$ follows $\tht_{1\uparrow}$ in the anticlockwise direction in $\cC_{\tht_1,\tht_2}$. To denote the same contour with opposite orientation we will use the superscript $``-"$, e.g., $\cC_{\tht_1,\tht_2}^-$.
		
		\begin{minipage}{\linewidth}
		\centering
		\begin{tabular}{lccr}
		\includegraphics[scale=.35]{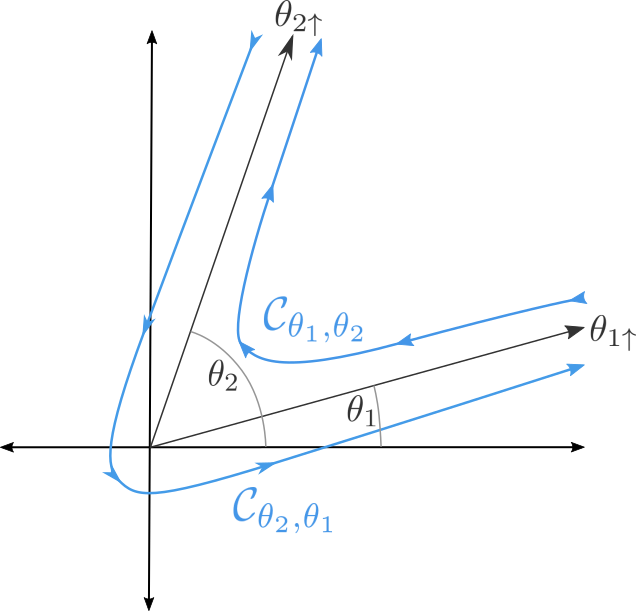} &&& \includegraphics[scale=.35]{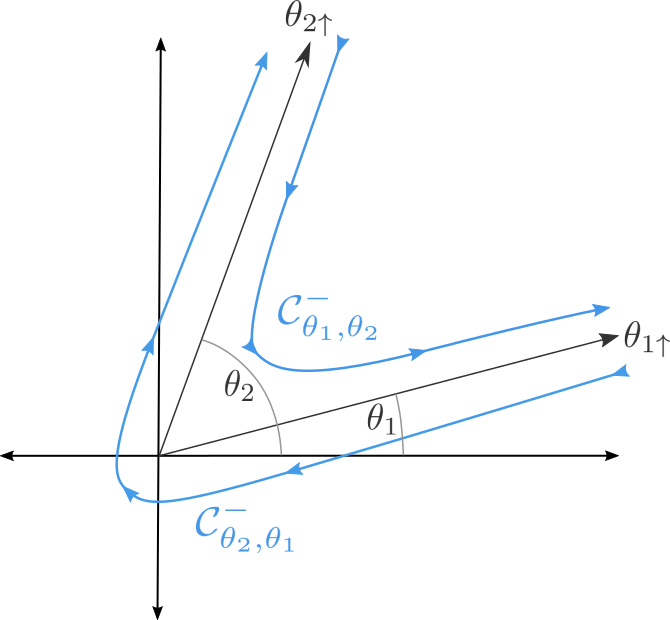}
		\end{tabular} \par
		Figure: Some exemplary contours.
		\end{minipage}
		\smallskip
		
		We define the following angles and cycles:
		\begin{align}
			\vartheta_a := \frac{\pi(2a - 1)}{k+2}\,, \qquad& \varphi_a := \frac{2\pi a}{k+2} \nn\\
			 \qquad C_a := \cC_{\vartheta_a, \vartheta_{a+1}}\,, \qquad& \wt C_a := \cC_{\varphi_a, \varphi_{a+1}}\,, \qquad \mbox{for} \quad a \in \bZ\,, \label{cycle1}
		\end{align}
		In the figure bellow we draw a couple of these cycles and point out their intersection numbers. If two cycles $C_a$ and $\wt C_b$ intersect at a point then we denote the intersection number of that point simply by $ C \circ \wt  C $ when the point being referred to is understood.\\
		\begin{minipage}{\linewidth}
			\centering
			\includegraphics[scale=.45]{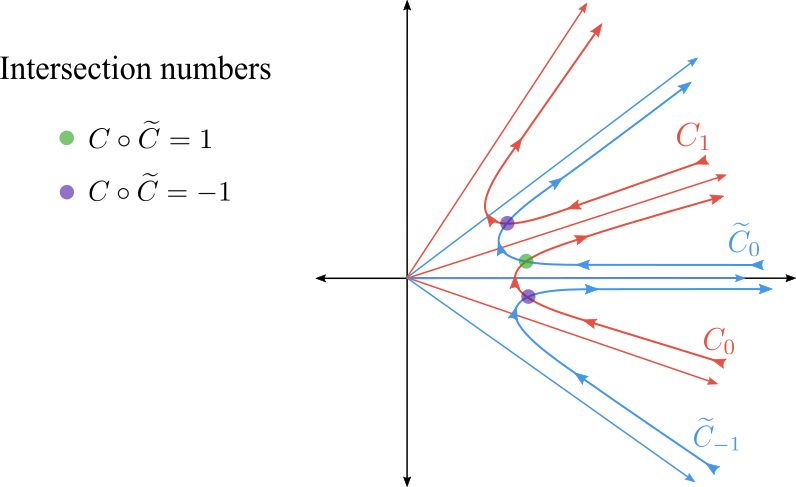}
			\par
			Figure: The cycles $C$ and $\wt C$.
		\end{minipage}
		\smallskip
		
		Now we have:
		\beq
			\oint_{C_a} \om_m  = e^{\frac{2\pi i a(m+1)}{k+2}} \oint_{C_0} \om_m\,, \qquad \oint_{\wt C_{-a-1}^-} \wt\om_n  = e^{-\frac{2\pi i a(n+1)}{k+2}} \oint_{\wt C_{-1}^-} \wt\om_n\,, \label{cshift}
		\eeq
		which follows after redefining integration variables as $X \to X e^\frac{2\pi i a}{k+2}$ and $\ov X \to \ov X e^{-\frac{2\pi i a}{k+2}}$ respectively. Note that if we set $X = re^{i\vartheta_a}$ and $\ov X = re^{i\varphi_a}$ with $r > 0$ then we get the following asymptotic behaviours for the one forms:
		\beq
			r \to \infty: \quad \om_m \sim r^m e^{i(m+1)\vartheta_a} e^{-r^{k+2}}\, \dd r\,, \qquad \wt\om_n \sim r^n e^{i(n+1)\varphi_a} e^{-r^{k+2}}\,\dd r\,,
		\eeq
		making the integrals well defined. We now proceed to evaluate $\oint_{C_0} \om_m$. We define a new variable:
		\beq
			Y := X^{k+2}\,
		\eeq
		in terms of which we can write:
		\beq
			X = Y^\frac{1}{k+2}\,, \qquad \dd X = \frac{1}{k+2} Y^\frac{-k-1}{k+2} \dd Y\,.
		\eeq
		and the angles defining the contour $C_0$ change as:
		\beq
			\vartheta_0 = -\frac{\pi}{k+2} \to -\pi\,, \qquad \vartheta_1 = \frac{\pi}{k+2} \to \pi\,.
		\eeq
		Since they represent the same direction in the complex plane we will write $\pi^\pm := \pi \pm \ep$ for the angles, where $\ep>0$ is infinitesimal, to keep track of the orientation of the resulting contour. Therefore,
		\beq
			\oint_{C_0} \om_m = \frac{1}{k+2} \oint_{\cC_{\pi^+,\pi^-}} Y^{-\frac{k-m+1}{k+2}} e^Y\, \dd Y
			= \frac{1}{k+2} \frac{2\pi i}{\Ga\lt(\frac{k-m+1}{k+2}\rt)}\,. \label{int1}
		\eeq
		To get the last equality we used the integral form of the reciprocal Gamma function:
		\beq
			\frac{1}{\Ga(z)} = \frac{1}{2\pi i} \oint_{\cC_{\pi^+,\pi^-}} \dd t\, t^{-z} e^t\,. \label{Grec}
		\eeq
		Similarly by defining $\ov Y = -\ov X^{k+2}$ we find:
		\beq
			\oint_{\wt C_{-1}^-} \wt\om_n = -\frac{e^{\frac{\pi i}{k+2}(n+1)}}{k+2} \oint_{\cC_{\pi^+,\pi^-}} \ov Y^{-\frac{k-n+1}{k+2}} e^{\ov Y} \dd \ov Y 
			= -2i\frac{e^{\frac{\pi i}{k+2}(n+1)}}{k+2} \sin\lt(\frac{\pi(n+1)}{k+2}\rt) \Ga\lt(\frac{n+1}{k+2}\rt) \,. \label{int2}
		\eeq
		The last equality is a combination of \rf{Grec} and Euler's reflection formula:
		\beq
			\Ga(1-z)\Ga(z) = \frac{\pi}{\sin(\pi z)}\,.
		\eeq
		
		The cycles defined in \rf{cycle1} are distinct for $a=0, \cdots,k+1$ and they satisfy:
		\beq
			\sum_{a=0}^{k+1} C_a = \sum_{a=0}^{k+1} \wt C_a = 0\,.
		\eeq
		The cycles $\wt C_a$ are dual to the cycles $C_a$ with the intersection form: 
		\beq
			 I_{ab} := C_a \circ C_b  = \de_{a,b} - \de_{a,b+1}\,, \label{intform}
		\eeq
		with inverse (restricting to $a,b=1,\cdots,k+1$ for independence):
		\beq
			I^{-1}_{ab} = \lt\{\begin{array}{cl} 1 & \mbox{when } a \ge b \\ 0 & \mbox{otherwise} \end{array} \rt.\,.
		\eeq
		Complex conjugation acts on the contours as follows:
		\beq
			\wt C_a^* = \cC_{\varphi_a, \varphi_{a+1}}^* = \cC_{-\varphi_{a+1}, -\varphi_a}^- = \cC_{\varphi_{-a-1}, \varphi_{-a}}^- = \wt C_{-a-1}^-
		\eeq
		Now, the generalization of Riemann bilinear identity \cite{Gaiotto:2013rk} gives us:
		\begin{align}
			\int_\bC \om_m \wedge \wt\om_n =&\; -\sum_{a,b=1}^{k+1} I^{-1}_{ab} \oint_{C_a} \om_m \oint_{\wt C_b^*} \wt \om_n
			= -  \sum_{a=1}^{k+1} \sum_{b=1}^a \oint_{C_a} \om_m \oint_{\wt C_{-b-1}^-} \wt \om_n \nn\\
			=&\; - \sum_{a=1}^{k+1} \sum_{b=1}^a e^{\frac{2\pi i}{k+2}[a(m+1)-b(n+1)]} \oint_{C_0} \om_m \oint_{\wt C_{-1}^-} \wt \om_n\,, \qquad \mbox{using \rf{cshift}}\,. \label{e1}
		\end{align}
		The integrals are independent of $a$ and $b$, so we can evaluate the sum separately:
		\beq
			\sum_{a=1}^{k+1} \sum_{b=1}^a e^{\frac{2\pi i}{k+2}[a(m+1)-b(n+1)]} = \frac{e^{-\frac{2\pi i}{k+2}(n+1)}}{1-e^{-\frac{2\pi i}{k+2}(n+1)}} \lt[\sum_{a=1}^{k+1} e^{\frac{2\pi ia}{k+2}(m+1)} - \sum_{a=1}^{k+1} e^{\frac{2\pi ia}{k+2}(m-n)} \rt] \label{e2}
		\eeq
		Note that for $m+1 \equiv 0\, (\mbox{mod}(k+2))$ \rf{int1} vanishes since the Gamma function in the denominator acquires a pole, and therefore the expression \rf{e1} will vanish as well. Assuming $m+1 \not\equiv 0\, (\mbox{mod}(k+2))$ we see that the first sum inside the parentheses in \rf{e2} is a sum over roots of unity excluding 1, therefore the first sum is $-1$. The second sum is also a sum over roots of unity excluding 1 unless $(m-n) \equiv 0\, (\mbox{mod}(k+2))$. Therefore, whenever $(m-n) \not\equiv 0\, (\mbox{mod}(k+2))$ the second sum is $-1$ and the expression \rf{e2}, and consequently \rf{e1}, vanish. From now on we assume that there exists a $q \in \bZ$ such that $m-n+q(k+2)=0$. Then \rf{e2} reduces to:
		\beq
			\sum_{a=1}^{k+1} \sum_{b=1}^a e^{\frac{2\pi i}{k+2}[a(m+1)-b(n+1)]} = \frac{- e^{-\frac{\pi i}{k+2}(n+1)}}{2i\sin\lt(\frac{\pi(n+1)}{k+2}\rt)} (k+2) \label{e3}
		\eeq
		Substituting \rf{e3}, \rf{int1} and \rf{int2} in \rf{e1} we find:
		\beq
			\frac{i}{2}\int_\bC \om_m \wedge \wt\om_n = \frac{\pi}{k+2} \frac{\Ga\lt(\frac{m+1}{k+2}+q\rt)}{\Ga\lt(\frac{k-m+1}{k+2}\rt)} \label{wmwn}
		\eeq
		Substituting this into \rf{XmXn2} and using \rf{XmXn1} we get the extremal correlators on $S^2$ \rf{Mmn}.

	\bibliography{note}
\end{document}